\documentclass[draftcls, 12pt, onecolumn]{IEEEtran}

\newcommand{\sss}{\scriptscriptstyle}
\newcommand{\mbs}[1]{\boldsymbol{#1}}
\renewcommand{\mbs}[1]{\mathbf{#1}}
\renewcommand{\mbs}[1]{\pmb{#1}}

\renewcommand{\th}{\textrm{th}}

\newcommand{\vect}[1]{{\lowercase{\mbs{#1}}}}
\newcommand{\mat}[1]{{\uppercase{\mbs{#1}}}}
\newcommand{\wt}{\widetilde}
\newcommand{\wh}{\widehat}

\usepackage{footnote} 
\usepackage{url}
\usepackage{cite}
\usepackage{mdwtab}
\usepackage{epsfig}
\usepackage{subfigure}
\usepackage{amsmath,amssymb,amsfonts,mathrsfs,amscd}%,amsthm}
\newtheorem{proposition}{Proposition}

\newtheorem{definition}{Definition}
\newtheorem{theorem}{Theorem}

\newtheorem{corollary}{Corollary}

\newtheorem{fact}{Fact}
\newtheorem{lemma}{Lemma}

\newcommand{\bb}[1]{\mathbb{#1}}
\newcommand{\nnb}{\nonumber}

\newcommand{\Fig}[1]{Fig.~\!\ref{#1}}

\newcommand{\Eq}[1]{(\ref{#1})}

\newcommand{\ie}{\emph{i.e.}}
\newcommand{\eg}{\emph{e.g.}}
\newcommand{\vs}{\emph{vs. }}

\newcommand{\etal}{\emph{et al.}\,}

\newcommand{\D}{\displaystyle}

\newcommand{\mysmallarraydecl}{\renewcommand{%
\IEEEeqnarraymathstyle}{\scriptstyle}%
\renewcommand{\baselinestretch}{1.1}    
\settowidth{\normalbaselineskip}{\scriptsize
\hspace{\baselinestretch\baselineskip}}%
\setlength{\baselineskip}{\normalbaselineskip}%
\setlength{\jot}{0.25\normalbaselineskip}%
\setlength{\arraycolsep}{2pt}}

\renewcommand{\matrix}[1]{\begin{bmatrix}#1\end{bmatrix}}

\newcommand\transcsymbol{\scriptscriptstyle \dag \!}
\newcommand\transsymbol{\scriptscriptstyle \mathsf{T} \!}

\newcommand\Abs[1]{\left|#1\right|}

\newcommand\Abssqr[1]{\left|#1\right|^2}

\newcommand{\inv}[1]{{#1}^{\scriptscriptstyle -1 \!}}
\newcommand{\transc}[1]{{#1}^{\transcsymbol}}
\newcommand{\trans}[1]{{#1}^{\transsymbol}}
\newcommand{\pstv}[1]{{#1}^{\sss +}}

\newcommand\diag{\mathrm{diag}}

\newcommand\Norm[1]{\left\|{#1}\right\|}
\newcommand\Frob[1]{\Norm{#1}^2_{\textrm{F}}}
\newcommand\Frobb[1]{\Norm{#1}_{\textrm{F}}}

\newcommand\defeq{\triangleq}

\newcommand{\Id}{\mathbf{I}}
\newcommand{\CN}[1][\Id]{\Ccal\Ncal\!\left(0,#1\right)}
\newcommand{\SNR}{{\mathsf{SNR}} }

\newcommand\iid{i.i.d.\ }

\newcommand\Pe{P_{\textrm{e}}}
\newcommand\Pout{P_{\textrm{out}}}

\newcommand\CC{\bb{C}}
\newcommand\EE{\bb{E}}

\newcommand\RR{\bb{R}}

\newcommand\Acal{\mathcal{A}}

\newcommand\Ccal{\mathcal{C}}

\newcommand\Ecal{\mathcal{E}}
\newcommand\Fcal{\mathcal{F}}

\newcommand\Ical{\mathcal{I}}

\newcommand\Ncal{\mathcal{N}}
\newcommand\Ocal{\mathcal{O}}

\newcommand\Scal{\mathcal{S}}
\newcommand\Tcal{\mathcal{T}}

\newcommand{\mA}{\mat{A}}
\newcommand{\mB}{\mat{B}}

\newcommand{\mG}{\mat{G}}
\newcommand{\mH}{\mat{H}}

\newcommand{\mL}{\mat{L}}
\newcommand{\mM}{\mat{M}}

\newcommand{\mT}{\mat{T}}
\newcommand{\mU}{\mat{U}}

\newcommand{\mX}{\mat{X}}
\newcommand{\mY}{\mat{Y}}
\newcommand{\mZ}{\mat{Z}}

\newcommand{\ma}{\vect{a}}

\newcommand{\mc}{\vect{c}}
\newcommand{\md}{\vect{d}}

\newcommand{\mf}{\vect{f}}

\newcommand{\mv}{\vect{v}}

\newcommand{\mx}{\vect{x}}
\newcommand{\my}{\vect{y}}
\newcommand{\mz}{\vect{z}}

\newcommand{\mGamma}{\mbs{\Gamma}}

\newcommand{\mSigma}{\mbs{\Sigma}}

\newcommand{\mlambda}{\boldsymbol{\lambda}}
\newcommand{\malpha}{\boldsymbol{\alpha}}

\newcommand{\NC}{\newcommand}
\newcommand{\RNC}{\renewcommand}

\NC{\He}{\mbs{\wt{H}}}
\NC{\mSigmasqrt}{\mGamma}
\NC{\mSigmasqrttrc}{\transc{\mSigmasqrt}}
\NC{\mSigmainv}{\mSigma^{-{1}}}
\NC{\dNAFMIMO}{d_{\NAF}^{\MIMO}}
\NC{\dNAFMIMON}{d_{\NAF,N}^{\MIMO}}
\NC{\AF}{\!\sss\Acal\Fcal}
\RNC{\AF}{\sss\textrm{AF}}

\NC{\Ixy}{\Ical(\mx;\sqrt{\SNR}\He\mx+\mz)}

\newcommand{\NAF}{\sss\textrm{NAF}}

\newcommand{\asympteq}{\doteq} %

\newcommand{\asymptleq}{\ \dot{\leq}\,}
\newcommand{\asymptgeq}{\ \dot{\geq}\,}

\newcommand{\Oe}{\Ocal}
\newcommand{\Oc}{\bar{\Ocal}}

\NC{\OneO}[2]{\One_{\Oe_{#1#2}}}
\NC{\Posr}{P_{\Oe_{sr}}}
\NC{\Poo}{P_{\Oe_{12},\Oe_{21}}}
\NC{\Pnoo}{P_{\bar{\Oe}_{12},\Oe_{21}}}
\NC{\Pono}{P_{\Oe_{12},\bar{\Oe}_{21}}}
\NC{\Pnono}{P_{\bar{\Oe}_{12},\bar{\Oe}_{21}}}
\NC{\out}{\textrm{out}}
\NC{\dHout}{d_{\mH}^{\out}}
\NC{\dout}{d_{\textrm{out}}}
\NC{\doutcc}{d^{\textrm{CC}}_{\textrm{out}}}
\NC{\doutnaf}{d^{\textrm{NAF}}_{\textrm{out}}}
\NC{\doutndf}{d^{\textrm{NDF}}_{\textrm{out}}}
\NC{\doutddf}{d^{\textrm{DDF}}_{\textrm{out}}}
\NC{\doutaf}{d^{\textrm{AF}}_{\textrm{out}}}
\NC{\doutdf}{d^{\textrm{DF}}_{\textrm{out}}}
\NC{\lmax}{\lambda_{\max}}
\NC{\Po}{P_{\Oe}}
\NC{\PE}{P_{\Ecal}}
\NC{\PEH}{P_{\Ecal_{\mH}}}
\NC{\PHH}{p_{\mH}(\mH)}
\NC{\PAA}{p_{\malpha}(\malpha)}
\NC{\dEmin}{d^2_{\min}}
\NC{\dEH}{d_{\Ecal|\mH}}

\NC{\Peo}{P_{\textrm{e},\Oe}}
\NC{\Peoc}{P_{\textrm{e},\Oc}}
\NC{\Pecondo}{P_{\textrm{e}|\Oe}}
\NC{\Pecondoc}{P_{\textrm{e}|\Oc}}
\NC{\Pepoc}{P_{\textrm{pe},\Oc}}
\NC{\Pepcondoc}{P_{\textrm{pe}|\Oc}}

\NC{\Tx}[1]{\textrm{Tx}(#1)}
\NC{\Rx}[1]{\textrm{Rx}(#1)}

\NC{\mHt}{\mbs{\wt H}}
\NC{\mHh}{\mbs{\wh H}}
\NC{\myt}{\mbs{\wt y}}
\NC{\mxt}{\mbs{\wt x}}
\NC{\mvt}{\mbs{\wt v}}
\NC{\mzetat}{\mbs{\wt \zeta}}
\NC{\mnt}{\mbs{\wt n}}
\NC{\mwt}{\mbs{\wt w}}
\NC{\mzt}{\mbs{\wt z}}

\NC{\range}[3]{\left\{#1\right\}^{#3}_{#2}}

\newcounter{mytempeqncnt}

\begin{document}

\title{Towards the Optimal Amplify-and-Forward Cooperative Diversity
  Scheme
 \thanks{Manuscript submitted to the IEEE
     Transactions on Information Theory.  The authors
     are with the Department of Communications and Electronics,
     \'{E}cole Nationale Sup\'{e}rieure des T\'{e}l\'{e}communications, 46, rue Barrault, 75013
     Paris, France~(e-mail: syang@enst.fr; belfiore@enst.fr).}
}
 \author{%   
 \authorblockN{Sheng Yang and Jean-Claude Belfiore}}%%\\

\maketitle
%\IEEEpeerreviewmaketitle

\begin{abstract}
  In a slow fading channel, how to find a cooperative diversity scheme
  that achieves the transmit diversity bound is still an open problem.
  In fact, all previously proposed amplify-and-forward~(AF) and
  decode-and-forward~(DF) schemes do not improve with the number of
  relays in terms of the diversity-multiplexing tradeoff~(DMT) for
  multiplexing gains $r$ higher than $0.5$. In this work, we study the
  class of slotted amplify-and-forward~(SAF) schemes. We first
  establish an upper bound on the DMT for any SAF scheme with an
  arbitrary number of relays $N$ and number of slots $M$.  Then, we
  propose a sequential SAF scheme that can exploit the potential
  diversity gain in the high multiplexing gain regime. More precisely,
  in certain conditions, the sequential SAF scheme achieves the
  proposed DMT upper bound which tends to the transmit diversity bound
  when $M$ goes to infinity. In particular, for the two-relay case,
  the three-slot sequential SAF scheme achieves the proposed upper
  bound and outperforms the two-relay non-orthorgonal
  amplify-and-forward~(NAF) scheme of Azarian \emph{et al.} for
  multiplexing gains $r\leq 2/3$. Numerical results reveal a
  significant gain of our scheme over the previously proposed AF
  schemes, especially in high spectral efficiency and large network
  size regime.
\end{abstract}


\renewcommand{\th}{\textrm{th}}
\NC{\HGA}{\mH_{\textrm{GA}}}
\NC{\HGAtrc}{\transc{\mH}_{\textrm{GA}}}

\section{Introduction and Problem Description}
\label{sec:intro}
As a new way to exploit spatial diversity in a wireless network,
cooperative diversity techniques have recently drawn more and more
attention.  Since the work of Sendonaris \emph{et
  al.}~\cite{Sendonaris1,Sendonaris2}, a flood of works has appeared
on this subject and many cooperative protocols have been
proposed~(see, for example,
\cite{LTW1,LTW2,Nabar,ElGamal_coop,Bletsas,Elia_relay}).  A
fundamental performance measure to evaluate different cooperative
schemes is the diversity-multiplexing tradeoff~(DMT) which was
introduced by Zheng and Tse~\cite{Zheng_Tse} for the MIMO Rayleigh
channel. It is well known that the DMT of any $N$-relay cooperative
diversity scheme is upper-bounded~(referred to as the \emph{transmit
  diversity bound} in \cite{LTW2}) by the DMT of a MISO system with
$N+1$ antennas,
\begin{equation}
  \label{eq:dmt_miso}
  d(r) = (N+1)\pstv{(1-r)}.
\end{equation}%
This bound is actually proved achievable by the cooperative multiple
access scheme~\cite{ElGamal_coop}, using a Gaussian code with an
infinite cooperation frame length.

However, how to achieve \Eq{eq:dmt_miso} in a single-user
setting~(\ie, half-duplex relay channel) in the general case is still
an open problem, even with an infinite cooperation frame length. In
the single-relay case, the best known cooperative scheme, in the class
of amplify-and-forward strategies, is the Non-orthogonal
Amplify-and-Forward (NAF) scheme and the Dynamic Decode-and-Forward
(DDF) scheme in the class of decode-and-forward strategies. The NAF
scheme was proposed by Nabar \etal~\cite{Nabar} and has been proved to
be the optimal amplify-and-forward scheme for a half-duplex
single-relay channel by Azarian \etal~\cite{ElGamal_coop}. It is
therefore impossible to achieve \Eq{eq:dmt_miso} by only
amplifying-and-forwarding with one relay. The DDF scheme was proposed
independently in \cite{ElGamal_coop, Tarokh_coop, Katz_coop} in
different contexts. In \cite{ElGamal_coop}, it is shown that the DDF
scheme does achieve \Eq{eq:dmt_miso} in the low multiplexing gain
regime~($r<0.5$) but it fails in the high multiplexing gain regime,
which is due to the \emph{causality} of the decode-and-forward scheme.
Intuitively, to achieve the MISO bound with a multiplexing gain $r$,
the source and the relay need to cooperate during at least $r$-portion
of the time.  However, before this might possibly happen, the relay
also needs at least $r$-portion of the time to decode the source
signal~(even with a Gaussian source-relay link). Therefore, it is
impossible for the DF schemes to achieve the MISO bound for $2r>1$.

\begin{figure}%[!htbp]
%\begin{minipage}{\textwidth}
  \begin{center}
    \centering\epsfig{figure=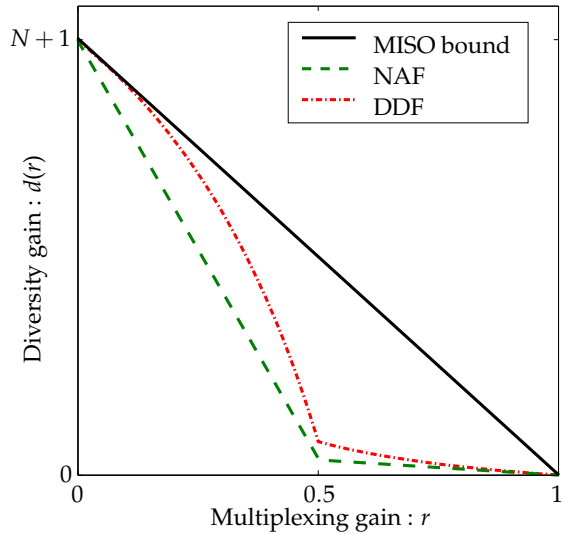,width=0.45\textwidth}
\caption{Diversity-multiplexing tradeoff of an $N$-relay channel~: NAF, DDF \vs MISO bound.}
\label{fig:dmt_miso_ddf_naf}      
  \end{center}
%\end{minipage}
\end{figure}

Being optimal in the single-relay case, the generalization of the NAF
and the DDF schemes proposed in \cite{ElGamal_coop}, also the best
known in each class, fails to exploit the potential spatial diversity
gain in the high multiplexing gain regime~($r>0.5$) with the growth of
the network size. The suboptimality of these two schemes becomes very
significant for a large number of relays, as shown in
Fig.~\ref{fig:dmt_miso_ddf_naf}.  Our goal is therefore to find a
practical scheme that can possibly fill the gap between the two
schemes and the MISO bound. In this work, we focus on the class of
slotted amplify-and-forward~(SAF) schemes because of the following
attractive properties~:
\begin{enumerate}
\item Low relaying complexity. The relays only need to scale the
  received signal and retransmit it.
\item Existence of optimal codes with finite framelength. We will show
  that any SAF scheme is equivalent to a linear fading channel,
  whose DMT is achieved by perfect \cite{Oggier-1} $M\times M$ codes.
  The code length for an $M$-slot SAF scheme is therefore at most
  $M^2$.
\item Flexibility. The source does not have to know the number of
  relays or the relaying procedure. The coding scheme only depends on
  the number of slots $M$ and is always optimal in terms of DMT.
\end{enumerate}

A natural question is raised~: \emph{Is it possible for a half-duplex SAF scheme
  to achieve the MISO bound~\Eq{eq:dmt_miso}? And how to achieve it if
  it is possible?} This question is partially answered in this work.
The main contributions of this work are as follows~:
\begin{itemize}
\item For a general $N$-relay $M$-slot SAF scheme, we establish a new
  upper bound~:
  \begin{equation}
    \label{eq:dmt_ub}
    d^*(r) = \pstv{(1-r)} + N\pstv{\left(1-\frac{M}{M-1}r\right)},
  \end{equation}%
  from which we conclude that it is impossible to achieve the MISO
  bound with a finite length, even without the half-duplex constraint.
  This bound is however tending to the MISO bound when $M$ goes to
  infinity. Then, we argue that the suboptimality of the $N$-relay NAF
  scheme is due to the fact that only half of the source signal is
  \emph{protected} by the relays.
\item Inspired by the upperbound \Eq{eq:dmt_ub}, we propose a
  half-duplex sequential SAF scheme. The basic idea is to let as many slots
  as possible~(\ie, $M-1$) be forwarded by the relays in the simplest
  way. For $M=2$ and an arbitrary $N$, the proposed scheme corresponds
  to the single-relay NAF scheme combined with the relay selection
  scheme~\cite{Bletsas} and the DMT upper bound is achieved. For
  arbitrary $(N,M)$, we show that the sequential SAF achieves the DMT
  upper bound in the extreme case where all relays are isolated from
  each other, \ie, there is no physical link between the relays.
  Nevertheless, even without the relays isolation assumption,
  simulation results show that a significant power gain over the NAF
  scheme is obtained by the sequential SAF scheme.
\item In particular, we show explicitly that the two-relay three-slot
  sequential SAF scheme dominates the two-relay NAF scheme for multiplexing
  gains $r\leq2/3$. It is therefore the best known two-relay
  amplify-and-forward scheme.
\end{itemize}

In this paper, we use boldface lower case letters $\mbs{v}$ to denote
vectors, boldface capital letters $\mbs{M}$ to denote matrices.
$\Ccal\Ncal$ represents the complex Gaussian random variable.
$\trans{[\cdot]},\transc{[\cdot]}$ respectively denote the matrix
transposition and conjugated transposition operations. $\Norm{\cdot}$
is the vector norm and $\Frobb{\cdot}$ is the Frobenius matrix norm.
$\pstv{(x)}$ means $\max(0,x)$. The dot equal operator $\asympteq$
denotes asymptotic equality in the high SNR regime, \ie,
\begin{equation*}
  p_1 \asympteq p_2 \quad   \textrm{means} \quad \lim_{{\sss\textsf{SNR}}\to\infty}\frac{\log p_1}{\log \SNR} = \lim_{{\sss\textsf{SNR}}\to\infty}\frac{\log p_2}{\log \SNR},
\end{equation*}%
and $\asymptleq,\asymptgeq$ are similarly defined.

The rest of the paper is organized as follows.
Section~\ref{sec:system-model} introduces the system model and the
class of SAF schemes. In Section~\ref{sec:ub}, we establish an
upper bound on the DMT of any SAF schemes, using a genie-aided model.
Then, Section~\ref{sec:sequential-SAF} proposes a sequential SAF scheme that
achieves the previously provided DMT upper bound in certain
conditions, when using two scheduling schemes. To show the performance
of the proposed scheme, numerical results with the sequential SAF scheme
are presented in Section~\ref{sec:results}, compared to the NAF scheme
and the non-cooperative scheme.  Finally, we provide some concluding
remarks in Section~\ref{sec:conclusion}.  For continuity of
demonstration, all detailed proofs are left in the Appendix.

\section{System Model}
\label{sec:system-model}

\subsection{Basic Assumptions}
\label{sec:basic-assumptions}

The considered system model consists of one source \textsf{s}, one
destination~\textsf{d} and $N$ relays~(cooperative terminals)
$\textsf{r}_1,\ldots,\textsf{r}_N$. The physical links between
terminals are slowly faded and are modeled as independent quasi-static
Rayleigh channels, \ie, the channel gains do not change during the
transmission of a cooperation frame, which is defined according to
different schemes~(protocols). The gain of the channel connecting
\textsf{s} and \textsf{d} is denoted by $g_0$. Similarly, $g_i$ and
$h_i$ respectively denote the channel gains between $\textsf{r}_i$ and
\textsf{d} and the ones between \textsf{s} and $\textsf{r}_i$.
$\gamma_{ij}$ is used to denote the channel gain between
$\textsf{r}_i$ and $\textsf{r}_j$. Channel quality between terminals
is parameterized by the variance of the channel gains. Unless
otherwise indicated, the relays work in half-duplex mode, that is,
they cannot transmit and receive at the same time.

\subsection{Slotted Amplify-and-Forward}
\label{sec:slott-ampl-forw}

\subsubsection{Definition}
\label{sec:definition}

In the paper, we study a particular class of amplify-and-forward
schemes that we call slotted amplify-and-forward~(SAF). More
precisely, an $N$-relay $M$-slot scheme is specified by the following
requirements~: 
\begin{itemize}
\item a cooperation frame is composed of $M$ slots of $l$ symbols,
  denoted by $\mx_i\in\CC^{l\times1},\ i=1,\ldots,M$;
\item during the $i^{\text{th}}$ slot, the source \textsf{s} transmits
  $\mx_i$ and the $j^{\text{th}}$ relay $\textsf{r}_j$, $j=1,\ldots,N$
  transmits $\mx_{r_j,i}\in\CC^{l\times1}$;
\item the received symbols at the $j^{\text{th}}$ relay and the
  destination are respectively denoted by $\my_{r_j,i},
  \my_i\in\CC^{l\times1}$, with
  \begin{equation}
    \label{eq:signal-model0}
    \left\{
      \begin{aligned}
        \my_{i}       &= g_0\,\mx_i + \sum_{j=1}^N g_{j}\,\mx_{r_j,i} + \mz_{d,i}  \\
        \my_{r_j,i} &= h_j\,\mx_i + \sum_{k=1,k\neq j}^N
        \gamma_{k,j}\,\mx_{r_k,i} + \mz_{r_j,i}
      \end{aligned}%
    \right.
  \end{equation}%
  where $\mz_{d,i},\mz_{r_j,i}\in\CC^{l\times1}$ are \iid AWGN with
  unit variance;
\item according to the AF constraint, $\mx_{r_j,i}$ can only be linear
  combination of the vectors $\my_{r_j,1},\ldots,\my_{r_j,i-1}$ that
  it receives in previous slots, \ie,
  \begin{equation}
    \label{eq:1}
    \mx_{r_j,i} = \sum_{k=1}^{i-1} p^{(j)}_{i,k} \my_{r_j,k}
  \end{equation}
  where $p^{(j)}_{i,k}$ depends on the AF protocol and the scheduling;
\item the transmitted signal $\mx_i$ and $\mx_{r_j,i}$ are subject to
  the short-term\footnote{We do not consider power control in our
    work.}  power constraint
  \begin{equation}
    \label{eq:PC}
      \EE\left(\Norm{\mx_i}^2+\sum_{j=1}^N\Norm{\mx_{r_j,i}}^2\right)\leq
  l\cdot\SNR,\quad \forall i.
  \end{equation}
%  where the expectation is taken on the
%  additive noises.
\end{itemize}
For example, the NAF scheme~\cite{ElGamal_coop} is an $N$-relay
$(2N)$-slot scheme and the non-orthogonal relay selection
scheme~\cite{Bletsas} is an $N$-relay two-slot scheme. Furthermore,
any AF scheme with cooperation frame length $L$ can also be regarded
as an $L$-slot SAF scheme with slot length constraint $l=1$.

In the SAF model, the knowledge of channel state information~(CSI) is
not specified. We assume that the cooperations between terminals are
coordinated by a scheduler~(that exists physically or logically).
Depending on how much CSI the scheduler has, the
coordination~(scheduling) can be static~(no CSI, \eg, NAF) or
dynamic~(based on global CSI, \eg, relay selection).  Therefore, for
each relay, the coefficients $\left\{p^{(j)}_{k,i}\right\}$'s in
\Eq{eq:1} are decided basing on its own CSI and the scheduling
information it receives from the scheduler. To be realistic, we assume
in our work that all terminals have receiver CSI only, and that
depending on applications the scheduler may have global CSI but can
only send order information to the relays, in order to minimize the
signaling overhead.

%% Since we restrict ourselves in the class of
%%   amplify-and-forward cooperation, the coordination information is
%%   simply \emph{what and when to send}.

%% the , is of length $M\,l$.  During any slot $i$,
%% $i=1,\ldots,M$, the source \textsf{s} transmits a sub-frame of $l$
%% symbols, denoted by a vector $\mx_i\in\CC^l$ and the relay
%% $\textsf{r}_j$, $j=1,\ldots,N$, can transmit $\mx_{r_j,i}\in\CC^l$, a
%% linear combination of the vectors it received in previous slots. Under
%% the half-duplex constraint, a relay does not receive while
%% transmitting. 

\subsubsection{Equivalent channel}
\label{sec:equivalent-channel}

Note that in the considered scheme, there is only one source signal
stream $[\mx_1\cdots\mx_M]$ and all relayed signal $\mx_{r_j,i}$ can
be eventually expressed as a noisy linear combination of
$\mx_1,\ldots,\mx_M$, as shown by \Eq{eq:signal-model0} and \Eq
{eq:1}. Therefore, without going to the details, we can verify that
the transmission of a cooperation frame with any SAF scheme described
above can be written in the following compact form
\begin{equation}
  \label{eq:vector_channel}
  \underbrace{\trans{[\my_1\cdots\my_M]}}_\mY = \sqrt{\SNR}\,\mH
  \underbrace{\trans{[\mx_1\cdots\mx_M]}}_\mX +
  \underbrace{\mZ_d+\mZ_e}_\mZ
\end{equation}
where $\mX\in\CC^{M\times l}$ is the normalized\footnote{For
  simplicity, we keep the same notation $\mx_1,\ldots,\mx_M$ to denote
  the normalized codeword.}~(by $\sqrt{\SNR}$) codeword matrix;
$\mH\in\CC^{M\times M}$ is the equivalent channel matrix consisting of
functions of the channel coefficients and the
$\left\{p^{(j)}_{i,k}\right\}$'s in \Eq{eq:1}; $\mZ_d\in\CC^{M\times
  l}\sim\CN[\Id]$ is the AWGN at the destination and
$\mZ_e\in\CC^{M\times l}\sim\CN[\mSigma_e]$ is the effective
accumulated noise\footnote{The $l$ columns of $\mZ_e$ are mutually
  independent and each column has the same covariance matrix
  $\mSigma_e$.} caused by the AF operations at each relay during the
whole transmission; the total noise is thus
$\mZ=\mZ_d+\mZ_e\sim\CN[\mSigma]$ with $\mSigma=\Id+\mSigma_e$.

%% Obviously, the transmission of a cooperation frame with any SAF scheme
%% is equivalent to $l$ channel uses of the following vector~(MIMO)
%% channel
%% \begin{equation}
%%   \label{eq:vector_channel}
%%   \my = \sqrt{\SNR}\,\mH \mx + \mz
%% \end{equation}%
%% where $\mx$ is the transmitted signal, $\mz\sim\CN[\mSigma_\mz]$ is
%% the equivalent additive colored noise with covariance matrix
%% $\mSigma_\mz$ and $\mH$ is an $M\times M$ lower-triangular matrix
%% representing the equivalent ``space-time'' channel between the source
%% and the destination. Moreover, we have $H_{ii}=c_i\,g_0$ with $c_i$
%% being a constant related to the transmission power.

\subsection{Diversity-Multiplexing Tradeoff and Achievability}
\label{sec:divers-mult-trad}

Let us recall the definition of the multiplexing and diversity gains.
\begin{definition}[Multiplexing and diversity gain\cite{Zheng_Tse}]
  A coding scheme $\{\Ccal(\SNR)\}$ is said to achieve
  \emph{multiplexing gain} $r$ and \emph{diversity gain} $d$ if
\begin{equation*}
  \lim_{{\sss\textsf{SNR}}\to\infty} \frac{R(\SNR)}{\log\SNR} = r \quad
\textrm{and}\quad 
  \lim_{{\sss\textsf{SNR}}\to\infty} \frac{\log\Pe(\SNR)}{\log\SNR} = -d
\end{equation*}
where $R(\SNR)$ is the data rate measured by bits per channel
use~(PCU) and $\Pe(\SNR)$ is the average error probability using the
maximum likelihood~(ML) decoder.
\end{definition}

\begin{theorem}\label{thm:dmt-saf}
  The DMT of any SAF scheme with equivalent channel model
  \Eq{eq:vector_channel} is
  \begin{equation}\label{eq:dmt-saf}
    d(r) = d_{\mH}(Mr),
  \end{equation}
  with $d_{\mH}(r)$ being the DMT of the linear channel
  \Eq{eq:vector_channel}. Furthermore, by vectorizing a full rate
  $M\times M$ space-time code with non-vanishing determinant~(NVD), we
  get a code that achieves the tradeoff $d(r)$ for the SAF scheme. The
  code construction only depends on the slot number $M$.
\end{theorem}
\begin{proof}
  The equality \Eq{eq:dmt-saf} is obvious, since $M$ is the
  normalization factor of the channel use. The achievability is
  immediate from the results in \cite{Tavildar,SY_JCB_coop}, stating
  that the DMT of a fading channel with any fading statistics can be
  achieved by a full rate NVD code.
\end{proof}

%% \subsection{Properties of the SAF Schemes}
%% \label{sec:prop-saf-schem}

%% Theorem~\ref{thm:dmt-saf} implies that for any SAF scheme, the optimal
%% code construction is available, using the NVD codes design~(see, for
%% example, \cite{Oggier-1, Elia_perfect}) and the code length is at
%% most\footnote{In some particular cases, the code length can be
%%   shorter. For example, the NAF scheme has a block-diagonal equivalent
%%   channel. As shown in \cite{SY_JCB_coop}, we can have an optimal code
%%   of length $2M$.}  $M^2$. 
Since the optimal code construction is independent of the fading
statistics of the channel, the only information that the source needs
for coding is the number of slots $M$. In practice, $M$ is decided by
the scheduler, based on the channel coherence time, decoding
complexity, etc. The relaying strategies are between the destination
and the relays and can be completely ignored by the source. When no
relay is helping, the equivalent channel matrix is diagonal. In this
case, even if the source is not aware of the non-relay situation, the
destination
can decode the signal with linear complexity. %%  when ``perfect'' codes are
%% used, since they are a rotated version of a vector of symbols from the
%% original constellation~($\ZZ^{Ml}$)\cite{Oggier-1, Elia_perfect}.  
All these properties make SAF schemes very flexible and suitable for
wireless networks, especially for \emph{ad hoc} networks where the
network topology changes frequently.

\section{Genie-Aided SAF and Upper Bound of the DMT}  
\label{sec:ub}
From \Eq{eq:signal-model0} and \Eq{eq:1}, it is clear that an SAF
scheme is actually defined by $\left\{p^{(j)}_{k,i}\right\}$. Therefore,
it is impossible to get the DMT of an SAF without precising
$\left\{p^{(j)}_{i,k}\right\}$. However, we can establish an upper bound
on the DMT of any SAF scheme, which is independent of the choice of
$\left\{p^{(j)}_{i,k}\right\}$. To this end, we will first introduce the
genie-aided SAF model.

\subsection{The Genie-Aided Model} 

We consider the following genie-aided model. We assume that before the
transmission of the $i^{\text{th}}$ slot, the relays know exactly the
coded signal $\mx_j$ for any $j<i$, via the genie. However, the relays
are not allowed to decode the message embedded in the signal, due to
the AF constraint. The half-duplex constraint is also relaxed.
Therefore, in the $i^{\text{th}}$ slot, the relays can transmit
\emph{any} linear combinations of the vectors
$\mx_1,\ldots,\mx_{i-1}$, \ie,
\begin{equation}
  \label{eq:GA-x}
  \mx_{r_j,i} = \sum_{k=1}^{i-1} l^{(j)}_{i,k} \mx_{k}
\end{equation}%
where $l^{(j)}_{i,k}$ can be set arbitrarily as long as the power
constraint \Eq{eq:PC} is satisfied. Obviously, the genie-aided SAF
provides better performance than the original SAF does, since unlike
in \Eq{eq:1} where we can only choose the coefficients of
$\my_{r_j,i}$, we are now free to choose the coefficients of $\mx_k$.
Moreover, there is no accumulated noise in the genie-aided model.

The equivalent channel model for the genie-aided SAF is still in the
form of \Eq{eq:vector_channel}, except that $\mZ_e=\mbs{0}$ and that
$\mH$ can be specified as
\begin{equation}
  \label{eq:GA-H}
  \mH = g_0\,\Id + \sum_{j=1}^N g_j\,\mL_j
\end{equation}%
where each matrix $\mL_j\in\CC^{M\times M}$ is strictly
lower-triangular with $\mL_j(i,k)=l^{(j)}_{i,k}$.

\subsection{Upper Bound on the DMT} 
\begin{theorem}\label{thm:theorem_UB}
  The optimal DMT of an $N$-relay $M$-slot genie-aided SAF scheme is
  \begin{equation}
    \label{eq:dmt_ub2}
    d^*(r) = \pstv{(1-r)} + N\pstv{\left(1-\frac{M}{M-1}r\right)},
  \end{equation}%
  for any $M>1$. It is achievable by using uniquely the relay with
  largest relay-destination gain to send $\mx_i$ in the $(i+1)^\th$
  slot, $i=1,\ldots,M-1$.
\end{theorem}
\begin{proof}
  See Appendix~\ref{app:proof_theorem_UB}.
\end{proof}
%% The following theorem states the best DMT that we can have with
%% genie-aided SAF schemes, which defines the upper bound of the DMT on
%% the original SAF schemes.
\begin{corollary}
  The DMT of any $N$-relay $M$-slot SAF scheme is upper-bounded by
  \Eq{eq:dmt_ub2} for any $M>1$.
\end{corollary}
In this theorem, we exclude the case $M=1$ for the obvious reason that
the single-slot SAF scheme corresponds to the non-cooperative case.
In the two-slot case~($M=2$), this upper bound is actually achievable
by previously proposed half-duplex schemes~: 1) with
single-relay~($N=1$), the NAF is shown in \cite{ElGamal_coop} to
achieve \Eq{eq:dmt_ub2}; 2) for $N>1$, the upper bound is achievable
by the relay selection NAF scheme~\cite{Bletsas} if the scheduler have
global CSI or by beamforming if the relays could have transmitter CSI.
Intuitively, the upper bound is tight in the two-slot case since the
half-duplex constraint is implicitly imposed by the SAF model.%%  And
%% the proposed scheme is actually the single-relay NAF scheme combined
%% with the relay selection scheme~\cite{Bletsas}.

On the other hand, in the single-relay case~($N=1$), the upper bound
is not tight for $M>2$~: it is shown in \cite{ElGamal_coop} that the
NAF scheme is the best single-relay half-duplex AF scheme in the DMT
sense. The looseness of the bound in the single-relay case is due to
the fact that the upper bound is obtained by relaxing the half-duplex
constraint which is too strong in the single-relay case.

\subsection{Implications}
\label{sec:discussion}

From the upper bound \Eq{eq:dmt_ub2}, two observations can be made ~:
1)~SAF schemes can never achieve the MISO bound with a finite number
of slots, even without the half-duplex constraint, and 2)~SAF schemes
can never beat the non-cooperative scheme for $r>\frac{M-1}{M}$. In
fact, the first observation can be seen as a necessary condition of
the second one, and it applies to all AF schemes as they can be seen
as $L$-slot SAF schemes.

Intuitively, even in the genie-aided model, the last slot is not
protected by any relay. This is due to the causality of the relay
channel, not to the half-duplex constraint. Therefore, at most $M-1$
slots out of $M$ slots can be protected, which explains the
suboptimality for $r>\frac{M-1}{M}$. In the same way, since only $N$
slots out of $2N$ slots are protected by one relay in the NAF scheme,
the NAF scheme is not better than the non-cooperative scheme for
$r>0.5$.

As stated in \cite{ElGamal_coop}, an important guideline for
cooperative diversity is to let the source keep transmitting all the
time so that the maximum multiplexing gain is achieved. Here, we
provide another guideline~: \emph{let most of the source signal be
  protected by extra paths}. Based on this guideline, we propose, in
next section, a sequential SAF scheme and we show that this scheme actually
achieves the upper bound \Eq{eq:dmt_ub2} in some particular cases.

\section{The Sequential SAF Scheme}  
\label{sec:sequential-SAF}

As previously stated, the NAF scheme is optimal in the single-relay
case, due to the half-duplex constraint. We consider the
multiple-relay case in the rest of the paper.

\begin{figure*}%[!htbp]
  \begin{minipage}{\textwidth}
  {{ \subfigure[NAF]
    {\includegraphics[height=0.2\textwidth]{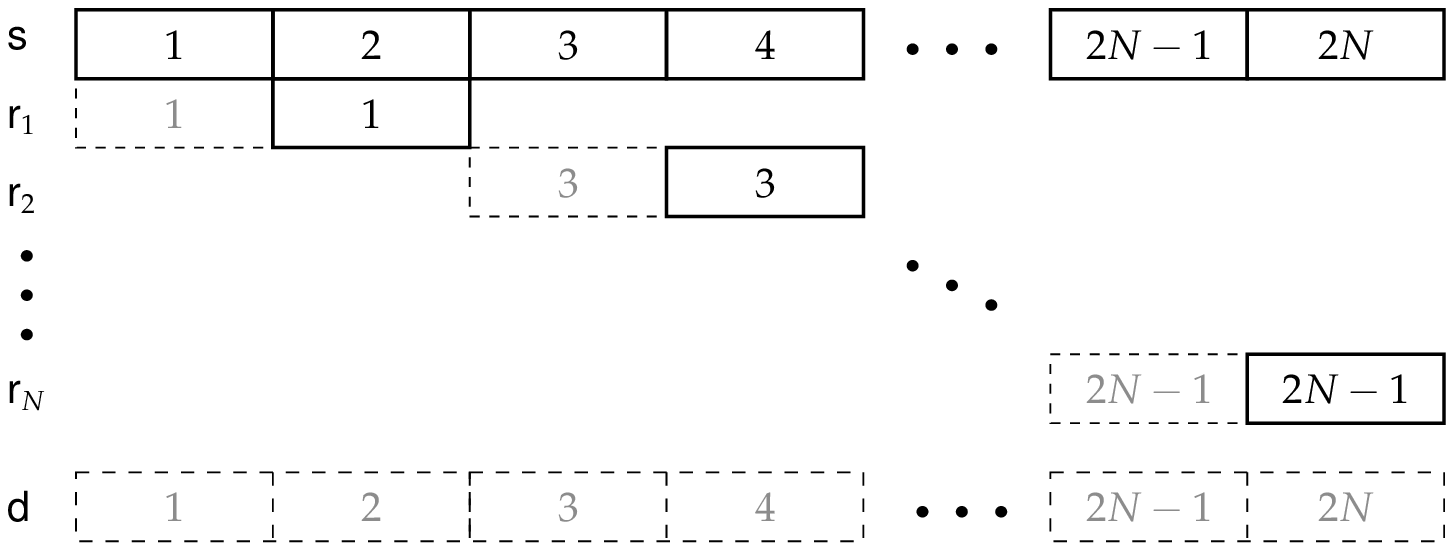}
\label{fig:frame-struct-NAF}}}
{\subfigure[Sequential SAF]
    {\includegraphics[height=0.2\textwidth]{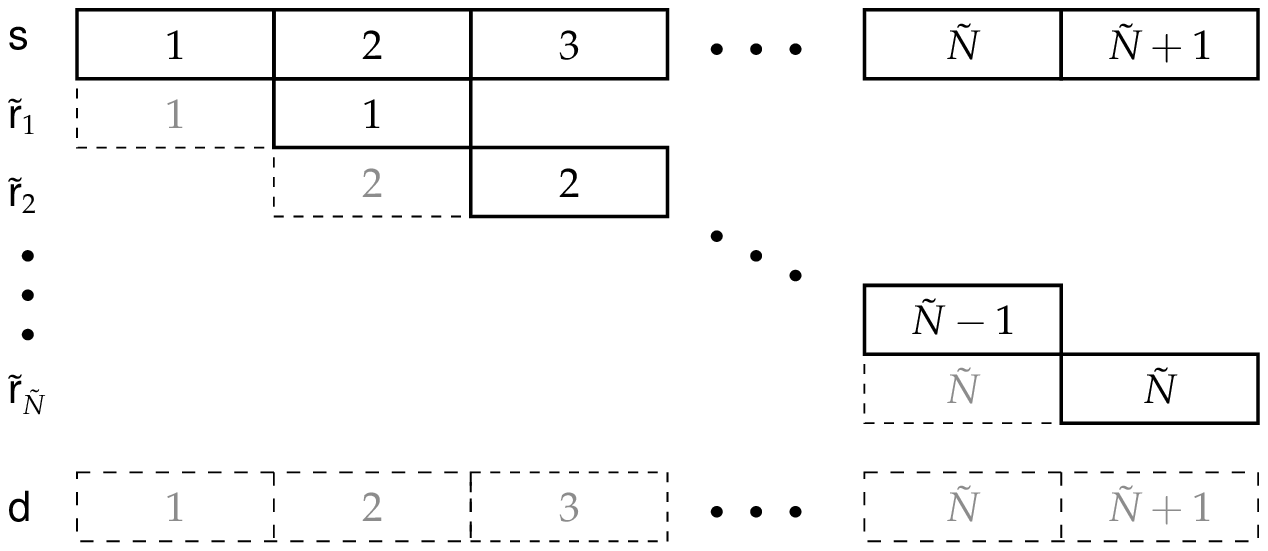}
\label{fig:frame-struct-SAF}}}}
\caption{Frame structure and relaying procedure of NAF and sequential SAF, solid box for transmitted signal and dashed box for received signal.}
\label{fig:frame-struct}
  \end{minipage}
\end{figure*}

Let us consider the following sequential SAF scheme. First of all, in order
to achieve the full multiplexing gain, the source must transmit during
all the $M$ slots.  Then, from the beginning of the second slot, in
each slot, there is one and only one relay forwarding a scaled version
of what it received in the previous slot. In such a way, $M-1$ slots
out of $M$ slots of the source signal are forwarded by at least one
relay. Here, we can see that this is only possible when we have more
than one relay, where different relays can alternatively help the
source to alleviate the half-duplex constraint. Thus, we have
$\tilde{N}\defeq M-1$ effective relays
$\tilde{\textsf{r}}_1,\ldots,\tilde{\textsf{r}}_{\tilde{N}}$ during
the transmission of a specific source.  The mapping between the real
relays and the effective relays is accomplished by relays scheduling
that will be discussed later on. The frame structure and the relaying
procedure are illustrated in Fig.~\ref{fig:frame-struct}, compared to
the NAF scheme.

\subsection{Equivalent Linear Fading Channel}
\label{sec:signal-model}

\newcommand{\gt}{\tilde{g}} \newcommand{\gmt}{\tilde{\gamma}}
\newcommand{\bt}{\tilde{b}} \newcommand{\htd}{\tilde{h}}
\newcommand{\bpi}{\bar{\pi}}

In SAF schemes, there is no difference in data processing for
different symbols within the same slot. Thus, we can consider one
symbol from a slot, without loss of generality. With the previous
description of the sequential SAF scheme, we have the following signal
model~:
\begin{equation}
\label{eq:signal-model1}
  \left\{
\begin{aligned}
  y_{d,i}   &= \sqrt{\pi_i\,\SNR}\,g_0\,x_i + \sqrt{\bpi_i\,\SNR}\,\gt_{i-1}\,\bt_{i-1}\,y_{r,i-1} + z_{d,i}  \\
  y_{r,i}   &= \sqrt{\pi_i\,\SNR}\,\htd_i\,x_i + \sqrt{\bpi_i\,\SNR}\,\gmt_{i-1,i}\,\bt_{i-1}\,y_{r,i-1} + z_{r,i}  
\end{aligned}%
\right.
\end{equation}%
%% \begin{equation}
%% \label{eq:signal-model1}
%%   \left\{
%% \begin{aligned}
%%   y_{d,i}   &= \sqrt{\pi_i\,\SNR}\,g_0\,x_i + \sqrt{\bpi_i\,\SNR}\,\gt_{i-1}\,\bt_{i-1}\,y_{r,i-1} + u_i  \\
%%   y_{r,i}   &= \sqrt{\pi_i\,\SNR}\,\htd_i\,x_i + \sqrt{\bpi_i\,\SNR}\,\gmt_{i-1,i}\,\bt_{i-1}\,y_{r,i-1} + v_i  
%% \end{aligned}%
%% \right.
%% \end{equation}%
where $x_i$ is the transmitted symbol from the source in the $i^\th$
slot; $y_{r,i}$ and $y_{d,i}$ are the received symbols at the $i^\th$
effective relay and at the destination, respectively, in the $i^\th$
slot; $\mz_{d,i}$'s and $\mz_{r,i}$'s are independent AWGN with unit
variance; $\htd_i$ and $\gt_i$, $i=1,\ldots,\tilde{N}$, are the
channel gains from the source to the $i^\th$ effective relay and from            
the $i^\th$ effective relay to the destination, respectively;
$\gmt_{i-1,i}$ is the channel gain between the $(i-1)^\th$ and the
$i^\th$ effective relay; $\bt_i$ is the processing gain at the $i^\th$
effective relay subject to the power constraint
$\EE{\left(|{\bt_i\,y_{r,i}}|^2\right)}\leq 1$. The power allocation
factors $\pi_i$, $\bpi_i$, $i=1,\ldots,M$ %% , are independent of the
%% channel coefficients and
satisfy $\sum_{i=1}^M(\pi_i+\bpi_i) = M.$ Finally, we set $\bpi_1=0$
and $\bt_0=0$.

%% By carefully treating the signal part and the noise part, 
We can express the signal model \Eq{eq:signal-model1} of $M$ slots in
the following vector form
\begin{equation}
\label{eq:signal-model_v}
  \left\{
\begin{aligned}
  \my_{d}   &= \sqrt{\SNR}\,g_0\,\diag(\ma)\,\mx + \mU_\mc\,\my_r + \mz_d \\
  \my_{r}   &=
  \sqrt{\SNR}\,\diag(\mbs{\tilde{h}})\,\diag(\ma)
  \mx + \mU_\md\,\my_r + \mz_r
\end{aligned}%
\right.
\end{equation}%
where $\mT\defeq \mU_\mc\inv{\left(\Id-\mU_\md\right)}$,
$\ma\in\RR_+^{M\times1}$ with $a_i\defeq\sqrt{\pi_i}$, and $\mU_\mc$,
$\mU_\md$ are $M\times M$ matrices defined as
\begin{align*}
  \mU_\mc &\defeq \matrix{\trans{\mbs{0}} & 0\\ \diag{(\mc)} & \mbs{0}} \\
  \mU_\md &\defeq \matrix{\trans{\mbs{0}} & 0\\ \diag{(\md)} & \mbs{0}} \\
\end{align*}%
with $\mc,\md\in\CC^{\tilde{N}\times1}$ whose components are defined
by $c_i\defeq\sqrt{\bpi_{i+1}\,\SNR}\,\gt_i\,\bt_i$ and
$d_i\defeq\sqrt{\bpi_{i+1}\,\SNR}\,\gmt_{i,i+1}\,\bt_i$ for
$i=1,\ldots,\tilde{N}$. Both $\mU_\mc$ and $\mU_\md$ are forward-shift
like matrices.

From \Eq{eq:signal-model_v}, we finally get the equivalent vector
channel
\begin{equation*}
  \my_d = \sqrt{\SNR}\,\mH\mx + \mz.
\end{equation*}%
where the equivalent channel matrix and noise are in the following
form~:
\begin{align}
  \mH &= \left(g_0\Id + \mT \diag(\mbs{\tilde{h}})\right) \diag(\ma) \label{eq:sequential-H}\\
  \mz &= \mz_d + \mT\,\mz_r,\label{eq:sequential-z}
\end{align}%
From \Eq{eq:sequential-z}, the covariance matrix of the noise is
$\mSigma_{\mz} = \Id+\mT\transc{\mT}$. We can show that the largest
and smallest eigenvalues of $\mSigma_\mz$ satisfy
$\lambda_{\max}\left(\mSigma_\mz\right) \asympteq
\lambda_{\min}\left(\mSigma_\mz\right) \asympteq \SNR^0$, which
implies that the DMT of the proposed scheme depends only on $\mH$ and
not on $\mSigma_\mz$.

Now, let us take a closer look at the equivalent channel matrix $\mH$,
which is lower-triangular. For simplicity, we ignore the term
$\diag(\ma)$ in our analysis since it does not impact the DMT. The
main diagonal of the equivalent channel is $g_0\,\Id$, representing
the direct~(source-destination) link. The off-diagonal entries are
defined by $\mT\,\diag(\mbs{\tilde{h}})$, where the $i^\th$
sub-diagonal\footnote{$\mT= \mU_\mc\,\inv{\left(\Id-\mU_\md\right)} =
  \mU_\mc\,\left(\Id+\mU_\md+\mU_\md^2 + \cdots\right)$.} is
$\mU_\mc\cdot\mU_\md^{i-1}\cdot\diag(\mbs{\tilde{h}})$, representing
the source-relays-destination $i$-hop link. Since the off-diagonal
entries are independent of the main diagonal entries, extra protection
to the source signal is provided and therefore the diversity gain is
obtained.

%% Unfortunately, the complex form of the equivalent channel matrix $\mH$
%% \Eq{eq:sequential-H} prevents us from obtaining the closed-form DMT in the
%% general case. Nevertheless, in some particular cases, we can have the
%% closed-form DMT and furthermore, it achieves the
%% upper bound~\Eq{eq:dmt_ub2}.

\subsection{Isolated Relays}
\label{sec:isolated-relays}

Calculating the DMT of the sequential SAF being prohibitive in general, we
search for an approximation. Intuitively speaking, the source signal
degrades with the number of hops, since the channel in each hop is
faded and that each normalization at the relays weakens the signal
power. Therefore, one possible approximation is to ignore the $i$-hop
links for $i>1$, which is equivalent to the special scenario where
relay $\textsf{r}_j$ is isolated with $\textsf{r}_{j-1}$ for
$j=2,\ldots,\tilde{N}$. In this case, the DMT can be obtained
explicitly. %We begin by the following proposition.
%% Let us consider a special scenario where
%% the relays have weak interconnections. In this case, we can assume
%% that the relays are isolated from each other. Then, the DMT of the
%% sequential SAF scheme can be obtained explicitly.
\begin{proposition} \label{prop:iso}
  When the relays are isolated from each other, \ie, $\gmt_{i,i+1}=0,\ 
  \forall i$, the DMT~\Eq{eq:dmt_ub2} is achievable with the sequential SAF
  scheme.
\end{proposition}
This proposition is proved in the following paragraphs. With the
assumption of relay isolation, we have $\mT=\mU_\mc$ and $\mH$ is
therefore a bidiagonal matrix. The special form of $\mH$ allows us get
the following lemma that is crucial to the proof.
\begin{lemma}
  \label{lemma:asymptgeq_D1}
  \begin{equation}    
%  \begin{multline}
\label{eq:asymptgeq_D1}
%    \begin{split}
    \max_{\mbs{\tilde{b}},\mbs{\pi},\mbs{\bar{\pi}}}\,\det\left(\Id +
      \SNR\mH\transc{\mH}\right) \asymptgeq
    \left(1+\SNR\Abssqr{g_0}\right)^M + \prod_{i=1}^{\tilde{N}}
    \left(1+\SNR \Abssqr{\gt_i\htd_i} \right)
%    \end{split}
%  \end{multline}%
  \end{equation}
\end{lemma}
\begin{proof}
  Using the bidiagonal property of $\mH$~(See
  Appendix~\ref{app:preliminaries} for details), we have 
  \begin{equation*}
%  \begin{multline*}%\label{eq:asymptgeq_D1}
%    \begin{split}
    \det\left(\Id +
      \SNR\mH\transc{\mH}\right) \geq
    \left(\SNR\Abssqr{g_0}\right)^M + \prod_{i=1}^{\tilde{N}}
    \left(1+\SNR \Abssqr{\gt_i\htd_i} \bar{\pi}_{i+1}\SNR \Abssqr{\tilde{b}_i} \right).
%    \end{split}
%  \end{multline*}%    
  \end{equation*}
  Since we can always find $\mbs{\pi}$,$\mbs{\bar{\pi}}$ and
  $\mbs{\tilde{b}}$ that satisfy simutaneously $\bar{\pi}_{i+1}\SNR
  |\tilde{b}_i|^2\asympteq \SNR^0$ and the power
  constraint~\Eq{eq:PC}, the lemma is proved.
\end{proof}
We can now introduce the scheduling strategies that permit the sequential
SAF to achieve the DMT upper bound~:
 \begin{enumerate}
 \item \emph{Dumb scheduling}: For $\tilde{N}=kN$ with $k$ being any
   integer, the relays help the source in a round-robin manner, \ie,
   $\tilde{\textsf{r}}_i=\textsf{r}_{(i-1)_N+1}$. For $\tilde{N}=kN+m$
   with $m\in[1,N-1]$, we first order the relays
   $\textsf{r}_1,\ldots,\textsf{r}_N$ in such a way that
   \begin{equation*}
%%      \min\{\Abs{g_1\,h_1},\ldots,\Abs{g_m\,h_m}\} \geq \max\{\Abs{g_{m+1}\,h_{m+1}},\ldots,\Abs{g_N\,h_N}\}.
     \min\{C_1,\ldots,C_m\} \geq \max\{C_{m+1},\ldots,C_N\}.
   \end{equation*}%
   where $C_i$ are the cost function defined by
   \begin{equation}
     \label{eq:costfunction}
     C_i \defeq \frac{\SNR^2\Abssqr{b_i\,g_i\,h_i}}{1+\SNR\Abssqr{b_i\,g_i}}.
   \end{equation}
   Then, we apply the round-robin scheduling. 
 \item \emph{Smart scheduling}: First, select the two ``best'' relays
   in the sense that they have largest cost function $C_i$ defined by
   \Eq{eq:costfunction}. Then, we apply the dumb scheduling on these
   two relays, as if we were in the two-relay $M$-slot case.
 \end{enumerate}
 
 These two scheduling strategies maximize \emph{statistically} the RHS
 of \Eq{eq:asymptgeq_D1} in the high SNR regime, so that upper bound
 \Eq{eq:dmt_ub2} is achieved. The detailed proof is provided in
 Appendix~\ref{app:proof_isolated}. Even though both schemes achieve
 DMT~\Eq{eq:dmt_ub2} under the relay isolation assumption, the smart
 scheme outperforms the dumb scheme in a general case, without relay
 isolation. Since the cost function $C_i$ is the effective SNR of the
 relayed signal at the destination if the $i^\th$ relay is used, the
 basic idea of the smart scheduling is to avoid using the ``bad''
 relays, where the noise level is higher than the other relays in
 average. Therefore, in $M$ slots, noise amplification is less
 significant with the smart scheduling than with the dumb scheduling.
 The impact is investigated in the next section, with the simulation
 results. Note that which scheduling scheme to be used depends
 strongly on the available CSI at the scheduler. If the scheduler has
 no CSI at all, dumb scheduling is used and we set $\tilde{N}=kN$~(or
 $M=kN+1$).
 
\begin{figure}%[!htbp]
%  \begin{minipage}{\textwidth}
    \begin{center}
      \epsfig{figure=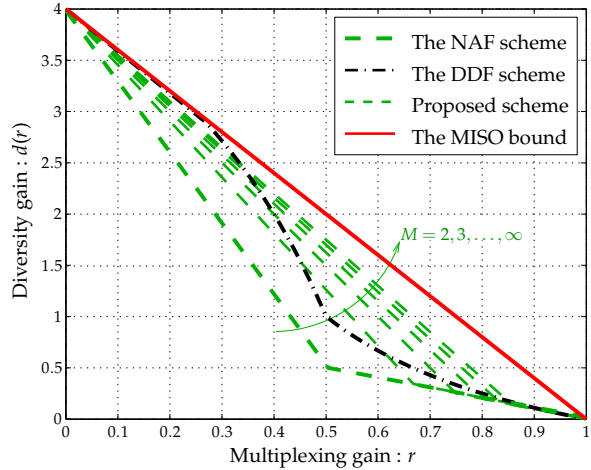,height=0.38\textwidth}
      \caption{Diversity-multiplexing tradeoff of different three-relay schemes with isolated relays.}
      \label{fig:dmt_comp_af_ddf_etc}      
    \end{center}
%  \end{minipage}
\end{figure}

 As an example, \Fig{fig:dmt_comp_af_ddf_etc} shows the DMT of
 different cooperative schemes for a three-relay channel, with relay
 isolation assumption. For $M=2$, the DMT of the proposed scheme
 coincides with that of the NAF scheme. With increasing $M$, the
 proposed scheme is approaching the MISO bound, which makes it
 asymptotically optimal.

\subsection{Non-Isolated Relays}
\label{sec:nonisolated-relays}

With interconnected relays, the DMT of the sequential SAF is generally
unknown, except for the following two cases.
\subsubsection{Two-Slot with Arbitrary Number of Relays}
\label{sec:two-slot-with}
 
\emph{Note that for the particular cases $M=2$, \ie, $k=0$ and $m=1$,
  the above analysis is valid whether the relays are isolated from
  each other or not.} This is because the maximum number of hops in
the channel is $1$. Therefore, the DMT \Eq{eq:dmt_ub2} for $M=2$ and
arbitrary $N$ is achieved by the sequential SAF with scheduler CSI, where
the scheduler selects the relay with largest $C_i$. It also
corresponds to the relay selection NAF scheme~\cite{Bletsas}.%%  if the
%% scheduler have global CSI or by beamforming if the relays could have
%% transmitter CSI. 
%% Intuitively, the upper bound is tight in the
%% two-slot case since the half-duplex constraint is implicitly imposed
%% by the SAF model.. And the proposed scheme is actually the
%% single-relay NAF scheme combined with the relay selection
%% scheme~\cite{Bletsas}.

\subsubsection{Two-Relay and Three-Slot}
\label{sec:two-relay-three}

\begin{figure}%[!b]
%  \begin{minipage}{\textwidth}
  \begin{center}
    \epsfig{figure=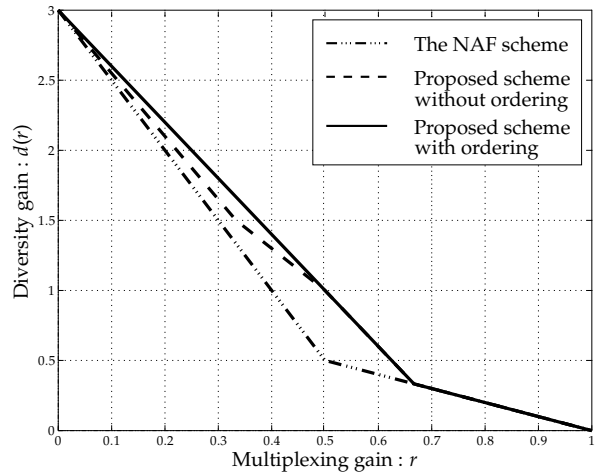,height=0.38\textwidth}
    \caption{Diversity-multiplexing tradeoff of the two-relay schemes.}
    \label{fig:dmt-proposed}        
  \end{center}
%  \end{minipage}
\end{figure}%

\begin{proposition}\label{prop:prop1}
  The two-relay three-slot sequential SAF scheme achieves the DMTs
  of \Fig{fig:dmt-proposed}, where the relay ordering is such that
  $\Abssqr{h_2}\geq\Abssqr{h_1}$, \ie, the relay with worse
  source-relay link transmits first.
\end{proposition}
\begin{proof}
  The DMTs are obtained with the same method as previously, by
  expressing explicitly the determinant
  $\det\left(\Id+\SNR\mH\transc{\mH}\right)$. See
  Appendix~\ref{app:proof-2r3s} for details.
\end{proof}
As shown in Appendix~\ref{app:proof-2r3s}, even though we have the
closed-form determinant expression, we can only have a lower-bound on
the DMT because of the complex determinant form. Unfortunately, the
lower-bound we get does not coincide with the
upper bound~\Eq{eq:dmt_ub2} for $r<0.5$. By adding a relay ordering
procedure~($\Abssqr{h_2}\geq\Abssqr{h_1}$), we finally get a
lower-bound equal to the upper bound. However, this does not
necessarily mean that the relay ordering improves the performance, as
we will show in the next section.

As shown in \Fig{fig:dmt-proposed}, the sequential SAF scheme~(with or
without relay ordering) outperforms the two-relay NAF scheme. Since
with the three-slot structure we protect $\frac{2}{3}$ of the source
signal, we can beat the non-cooperative scheme for $0\leq r \leq
\frac{2}{3}$.  It is therefore the best AF scheme known for the
two-relay case. To further improve the DMT, we should increase the
number of slots.

\subsection{Discussions}
\subsubsection{Artificial Relay-Isolation}
\begin{figure}%[!htbp]
%  \begin{minipage}{\textwidth}
    \begin{center}
      \epsfig{figure=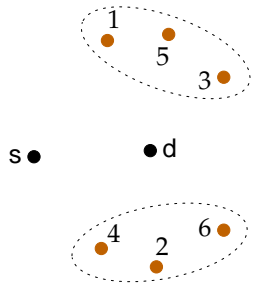,width=0.2\textwidth}
      \caption{A scheme to create weak inter-relay connections, in order to approximate the relay-isolation condition. The
        order of the relays are indicated by the numbers.}
      \label{fig:isolation}      
    \end{center}
%  \end{minipage}
\end{figure}%

Although it is hard to tell if the multi-hop links are harmful,
proposition~\ref{prop:iso} shows that the relay-isolation condition is
sufficient to achieve the DMT~\Eq{eq:dmt_ub2}. If the scheduler has
global CSI, it can order the relays in such a way that consecutive
relays are separated as far as possible to approximate the
relay-isolation condition. An example scheme is shown in
\Fig{fig:isolation}.

\subsubsection{Practical Considerations}
\label{sec:pract-cons}

%% To implement the sequential SAF schemes, the relay ordering is essential
%% for the smart scheduling and the $\tilde{N}\neq kN$ case of the dump
%% scheduling. An intelligent way to implement the relay ordering is to
%% employ a RTS/CTS like protocol as proposed in \cite{Bletsas}. In this
%% case, the relays measure the source-relay channel quality $\Abs{h_i}$
%% by the reception of the RTS frame from the source. Then, the
%% destination broadcasts a relay-probing frame, from which the relays
%% can estimate the relay-destination channel~$\Abs{g_i}$~(we assume a
%% TDD relay-destination link). The relay with the strongest product gain
%% $\Abs{g_i h_i}$ reacts first and becomes relay~1, and so on. Based on
%% the order, the destination decides a scheduling scheme and broadcasts
%% the parameters~(\eg, the relay ordering for the relays and number of
%% slots $M$ for the source, etc.) in the CTS frame. Since we only
%% consider slow fading channels, the ordering would not be so frequent
%% and the signaling overhead is negligible. In the worst case where the
%% above signaling is impossible, an order of cooperation for the relays
%% should be predefined and we apply the dump scheduling with a slot
%% number $M$ such that $M-1= kN$. In this case, the same DMT is
%% achieved.

In practice, an individual scheduler might not exist physically in the
network. In this case, we can integrate the scheduler's role into the
destination receiver. To implement the relay ordering, which is
essential for the smart scheduling and the $\tilde{N}\neq kN$ case of
the dumb scheduling, an intelligent way is similar to the RTS/CTS
scheme proposed in \cite{Bletsas} described as follows~:
\begin{itemize}
\item If we have the reciprocity for the forward and the backward
  relay-destination links, \ie, the channel gains are the same~($g_i$)
  for the forward and backward links, an intelligent way to implement
  the relay ordering is similar to the RTS/CTS scheme proposed in
  \cite{Bletsas}. First, the relays measure the source-relay channel
  quality $\Abs{h_i}$ by the reception of the
  \emph{RTS}~(Ready-to-Send) frame from the source.  Then, the
  destination broadcasts a \emph{relay-probing} frame, from which the
  relays can estimate the relay-destination channel~$\Abs{g_i}$. Each
  relay calculates the cost function $C_i$ and reacts by sending an
  \emph{availability} frame after $t_i$ time which is proportional to
  $C_i$. Therefore, the relay with the largest cost function is
  identified as relay~1, and so on.  Finally, based on the order, the
  destination decides a scheduling strategy and broadcasts the
  parameters~(\eg, the relay ordering for the relays and number of
  slots $M$ for the source, etc...) in the \emph{CTS}~(Clear-to-Send)
  frame.
\item When there is no reciprocity for the relay-destination links, we
  modify the last three steps as follows. Each relay quantizes the
  source-relay gain and sends it in the \emph{availability} frame to
  the destination using its own signature. Then, the destination can
  estimate the relay-destination links quality $\Abs{g_i}$ and also
  gets the estimates $\Abs{h_i}$ by decoding the signal. Finally, the
  destination decides the order based on the cost functions and
  broadcasts the \emph{CTS} frame.
\end{itemize}

Since we only consider slow fading channels, the ordering would not be
so frequent and the signaling overhead is negligible in both
cases~(the overhead issue is mentioned in \cite{Bletsas}). In the
worst case where the above signaling is impossible, a cooperation
order for the relays should be predefined and we apply the dumb
scheduling with a slot number $M$ such that $M-1= kN$.  In this case,
the same DMT is achieved.

\section{Numerical Results}
\label{sec:results}
In this section, we investigate the numerical results obtained by
Monte-Carlo simulations. By default, we consider a symmetric network,
where all the channel coefficients are i.i.d. Rayleigh distributed
with unit variance. There is therefore no \emph{a priori} advantage of
the source-relay links over the source-destination link. The power
allocation factors are $\pi_i=\bpi_i=0.5$ for $i=2,\ldots,M$ and
$\pi_1=1$. Information rate is measured in bits per channel
use~(BPCU). We compare the proposed sequential SAF scheme to the NAF scheme
and the non-cooperative scheme in both small network scenarios~($2$
relays) and large network scenarios~($12$ relays).

\subsection{Two-Relay Scenario}
\label{sec:two-relay-scenario}

\subsubsection{Three-Slot Case}
\label{sec:three-slot-case}

Fig.~\ref{fig:outage_se} shows the performance of the proposed
two-relay three-slot scheme for different spectral efficiencies. Note
that with a low spectral efficiency~($2$~BPCU), the proposed schemes
have almost the same performance as the NAF scheme. However, when
increasing the spectral efficiency, the gain of our schemes compared
to the NAF strategy increases. For $10$~BPCU, the NAF scheme barely
beats the non-cooperative scheme. Also note that in all cases, the
scheme with relay ordering proposed in Sec.~\ref{sec:two-relay-three}
is not better than the one without relay ordering. Based on that
observation, we conjecture that we can achieve the DMT~\Eq{eq:dmt_ub}
even without relay ordering in the two-relay three-slot case.

\newlength{\figwidth}
\setlength{\figwidth}{0.48\textwidth}
\begin{figure}%[!htbp]
%  \begin{minipage}{\textwidth}
    \begin{center}
      \epsfig{figure=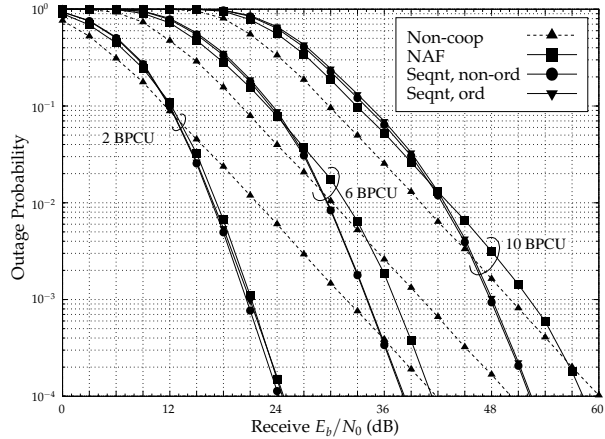,width=\figwidth}
      \caption{Outage probabilities for the non-cooperative, NAF and sequential SAF 
        scheme with three slots. Two-relay symmetric network.
        Considered information rates: $2$, $6$ and $10$ BPCU.}
      \label{fig:outage_se}      
    \end{center}
%  \end{minipage}
\end{figure}%

\begin{figure*}%[!htbp]
  \begin{minipage}{\textwidth}
  {{ \subfigure[Frame Error Rate]
    {\includegraphics[width=\figwidth]{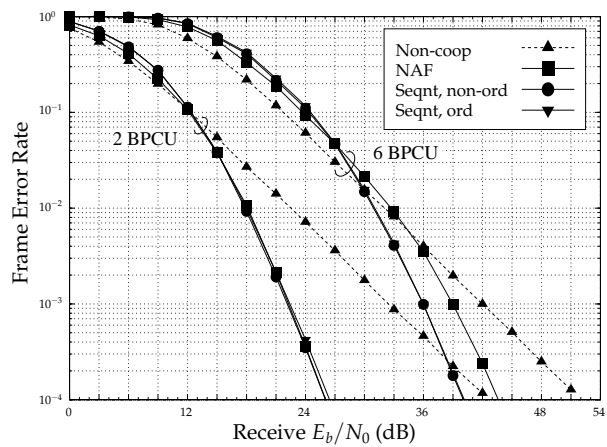}
\label{fig:2r3s_WER}}}
{\subfigure[Symbol Error Rate]
    {\includegraphics[width=\figwidth]{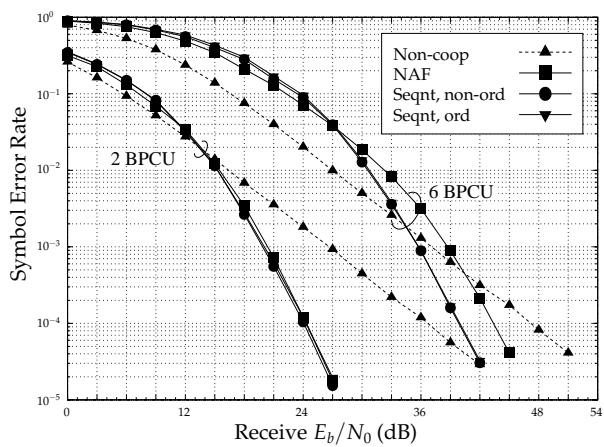}
\label{fig:2r3s_SER}}}}
\caption{Error  rate performance: sequential SAF \vs NAF scheme. 
  Two-relay symmetric network, perfect $3\times3$ code for the
  three-slot SAF scheme and $\Ccal_{2,1}$ for the NAF scheme for the
  NAF. $4$- and $64$-QAM for $2$ and $6$ BPCU, respectively.}
\label{fig:error-rate}
  \end{minipage}
\end{figure*}

\begin{figure*}%[!htbp]
  \begin{minipage}{0.5\textwidth}
%    \begin{center}
      \epsfig{figure=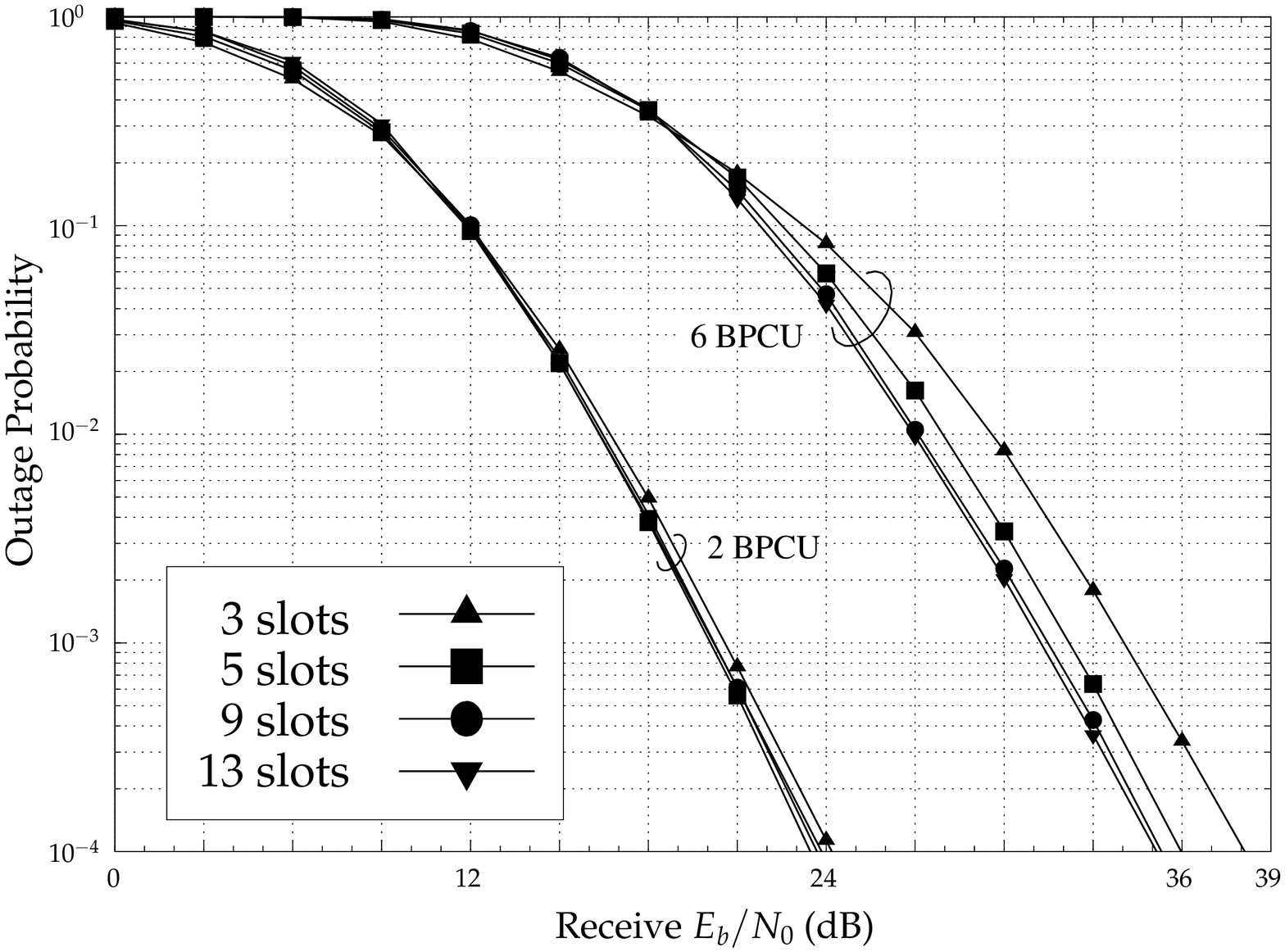,width=\figwidth}
      \caption{Outage probability of the sequential SAF scheme with $3$, $5$, $9$ and $13$ slots. 
        Two-relay symmetric network.}
      \label{fig:outage_sn}      
%    \end{center}
  \end{minipage}
%\end{figure}%
%\begin{figure}%[!htbp]
  \begin{minipage}{0.5\textwidth}
%    \begin{center}
      \epsfig{figure=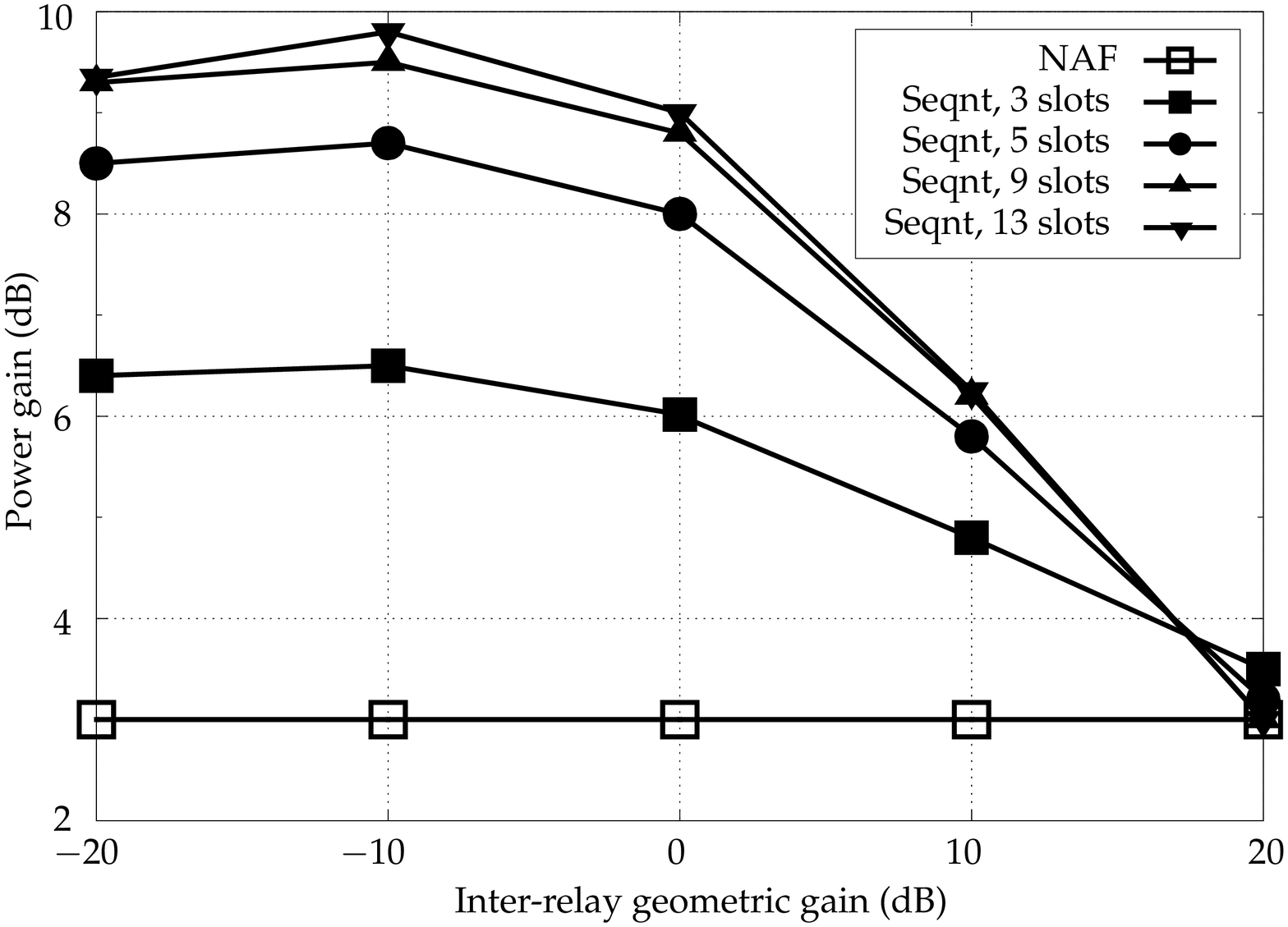,width=\figwidth}
      \caption{Power gain to the non-cooperative scheme~: impact of the inter-relay geometric gain. 
        Two-relay network. Target information rate~: $6$~BPCU. Target
        outage probability~: $10^{-3}$.}
      \label{fig:gain_10m3}      
%    \end{center}
  \end{minipage}
\end{figure*}%

\begin{figure*}%[!htbp]
%  \begin{minipage}{0.5\textwidth}
  \begin{center}
    \epsfig{figure=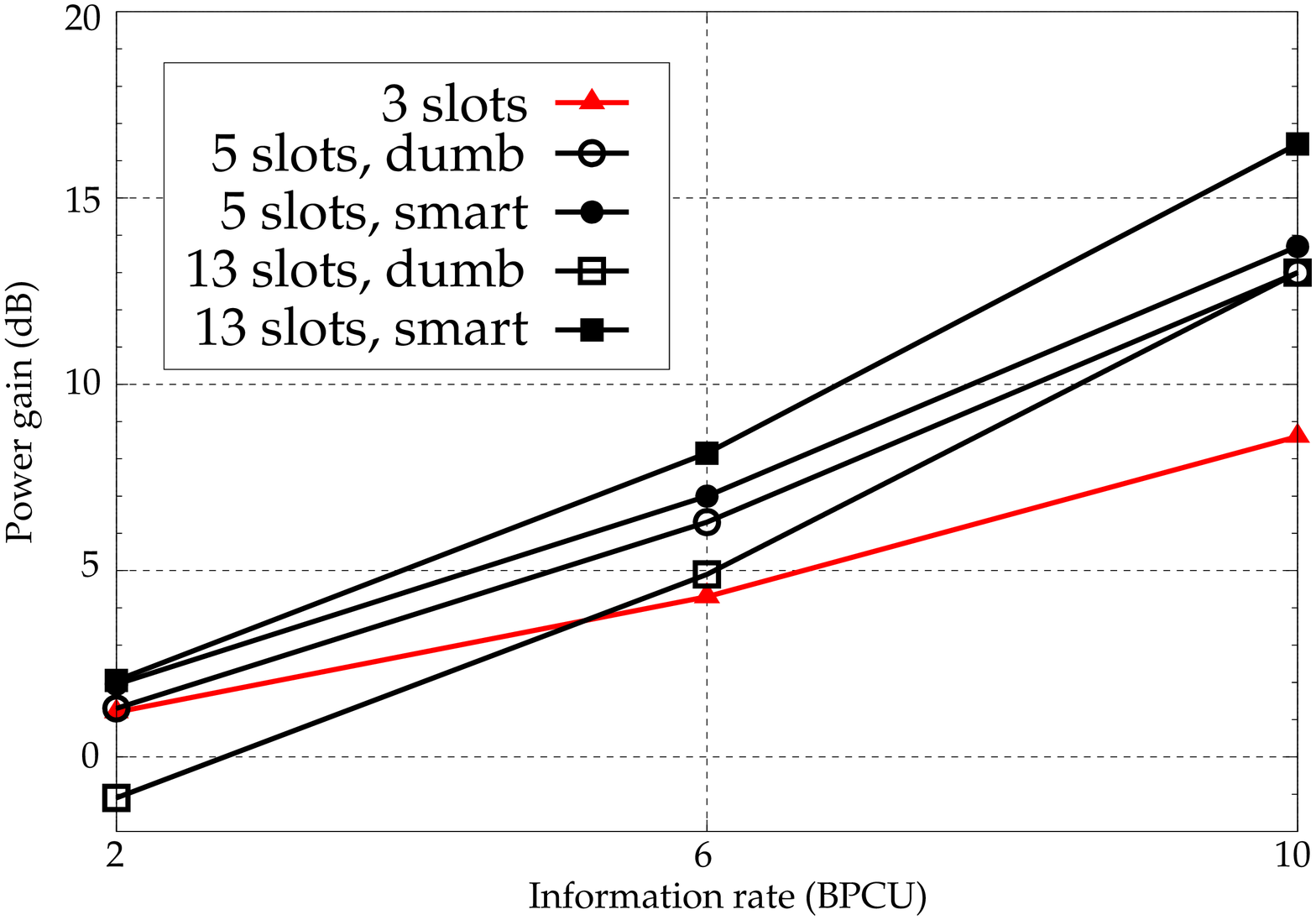,width=1.2\figwidth}
    \caption{Power gain to the NAF scheme with selection~: dumb \vs
      smart scheduling. Symmetric network with $12$ relays. Target
      outage probability~: $10^{-3}$. }  \label{fig:dump_vs_smart}    
  \end{center}
%  \end{minipage} 
\end{figure*}%

Then, we consider the error rate performance of NVD codes (\ie,
achieving the DMT) under ML decoding. For the two-relay NAF scheme, we
use the optimal code $\Ccal_{2,1}$~(QAM) proposed in
\cite{SY_JCB_coop}. For the sequential SAF scheme, we use the perfect
$3\times3$ code construction proposed in \cite{Elia_perfect}, based on
QAM constellations, the best known $3\times3$ real
rotation~\cite{Viterbo_table} and the ``non-norm'' element
$\gamma=\frac{1+2i}{2+i}$. The vectorized code~(frame) lengths are $8$
and $9$ QAM symbols for the NAF and the sequential SAF, respectively.
$4$-QAM and $64$-QAM uncoded constellations are used, corresponding to
the $2$ BPCU and $6$ BPCU counterparts in the outage performance. The
frame error rate~(FER) is shown in Fig.~\ref{fig:2r3s_WER}. It is
surprising to see such a similarity between code performance and
outage performance: for a given probability (error or outage
respectively), all SNR differences between the compared schemes are
almost the same. We have a power gain of more than $3$~dB for FER
lower than $10^{-3}$ with $64$-QAM.  For fairness of comparison
between different frame length, we also show the symbol error rate
performance in Fig.~\ref{fig:2r3s_SER}.

As stated in theorem~\ref{thm:dmt-saf}, we can always construct
optimal codes for a given SAF scheme. To focus on the cooperative
scheme itself, we only consider the outage probability hereafter.

\subsubsection{Impact of the Number of Slots}
\label{sec:slot-number}

Fig.~\ref{fig:outage_sn} shows the outage performance with different
numbers of slots. For $2$~BPCU, the difference is minor~(within
$1$~dB).  However, for $6$~BPCU, the power gain compared to the
three-slot scheme increases to $2$ and $3$~dB for $5$ slots and $13$
slots, respectively. The increasing SNR gain shows the superiority of
the schemes with a larger number of slots in terms of DMT, even without
the relay isolation assumption.

\subsubsection{Inter-Relay Geometric Gain}
\label{sec:inter-relay-geom}

In Fig.~\ref{fig:gain_10m3}, we show the impact of the inter-relay
geometric gain~(defined as $\EE\Abssqr{\gamma_{ij}}/\EE\Abssqr{h_j}$)
on the outage performance. In this scenario, all paths have the same
average channel gain~($0$~dB), except for the inter-relay channels
whose channel gains vary form $-20$~dB~(weak interconnection) to
$20$~dB~(strong interconnection). The y-axis represents the power gain
to the non-cooperative scheme with $6$ BPCU and outage probability of
$10^{-3}$. The x-axis represents the inter-relay geometric gain. As
shown in Fig.~\ref{fig:gain_10m3}, the NAF scheme is independent of
the geometric gain since there is no inter-relay communication at all
in the NAF scheme. In the weak interconnection regime~($<0$~dB), the
sequential SAF scheme is not sensitive to the geometric gain and we always
have a better performance by increasing the slot number. However, in
the strong interconnection regime~($>0$~dB), the performance degrades
dramatically with the increase of inter-relay gain and the increase of
the number of slots. Intuitively, the task of the $i^\th$ effective
relay is to protect the source signal $\mx_i$, transmitted in the
$i^\th$ slot. A strong interconnection between the $(i-1)^\th$ relay
and the $i^\th$ relay makes $\mx_i$ drowned in the combined signal of
$\mx_1,\ldots,\mx_{i-1}$ from the $(i-1)^\th$ relay.
%% \begin{figure}%[!htbp]
%% %  \begin{minipage}{\textwidth}
%%     \begin{center}
%%       \epsfig{figure=./figs/gain_10m3.eps,width=\figwidth}
%%       \caption{Power gain to the non-cooperative scheme~: impact of the inter-relay geometric gain. 
%%         Two-relay network. Target information rate~: $6$~BPCU. Target
%%         outage probability~: $10^{-3}$.}
%%       \label{fig:gain_10m3}      
%%     \end{center}
%% %  \end{minipage}
%% \end{figure}%

\subsection{Large Network~: Dumb \vs Smart Scheduling}
\label{sec:large-network}

Now, we consider a large symmetric network with $12$ available relays.
We compare the proposed scheme to the NAF scheme. To ensure fairness,
the considered NAF is combined with the relay selection scheme, \ie,
the source is only helped by the best relay~(with largest $C_i$). For
the sequential SAF scheme, both the dumb and the smart schedulings are
considered. Obviously, with $3$ slots, the dumb scheduling is the same
as the smart scheduling. As shown in Fig.~\ref{fig:dump_vs_smart}, the
power gain increases with spectral efficiency, showing the superiority
of our scheme in terms of DMT. The increase is more significant with a
larger slot number. With the same slot number, the curve of the dumb
scheduling is parallel to that of the smart scheduling, meaning the
same DMT for the same slot number.  The power gain is up to $8$ and
$16$ dB for $6$ BPCU and $10$ BPCU, respectively. For $2$ BPCU, the
13-slot dumb scheduling scheme is worse than the NAF, since the noise
amplification is significant. As we see, the smart scheduling is
always better than the dumb scheduling. In the considered cases, the
$5$-slot smart scheduling outperforms the $13$-slot dumb scheduling.
Since the optimal codes are respectively of length $5^2$ and $13^2$
for the $5$ slot and the $13$ slot cases, the use of smart scheduling
can dramatically reduce the decoding complexity.

\section{Conclusion and Future Work} 
\label{sec:conclusion}
In this paper, we considered the class of slotted amplify-and-forward
schemes. We first derived, for the SAF schemes, an upper bound of the
DMT which asymptotically~(when the framelength grows to infinity)
achieves the MISO bound. Then, we proposed and analyzed a sequential SAF
scheme for which the DMT upper bound is achieved in some special
cases. In particular, the two-relay three-slot sequential SAF is optimal
within the $N=2, M=3$ class and therefore outperforms all previously
proposed two-relay AF schemes.

The superiority of the sequential SAF scheme over the previously proposed
AF schemes lies in the fact that it exploits the potential diversity
gain in the high multiplexing gain regime~($r>0.5$), whereas all
previously proposed AF schemes do not beat the non-cooperative scheme
for $r>0.5$. An important guideline for the design of AF schemes was then proposed~: 
let most of the source signal be protected by extra
paths. We also showed that, by using a smart relay scheduling, the
complexity of decoding can be dramatically reduced. Numerical
results on both the outage and error rate performance reveal a
significant gain of our scheme compared to previously proposed AF schemes.
Since we can always find optimal codes of finite length for any SAF
scheme and the code construction is independent of the number of
relays, the proposed scheme is a combination of efficiency and
flexibility.

Even though we showed that the sequential SAF scheme is asymptotically
optimal in some particular cases, the DMT for the general case is
unknown. It would also be interesting to find a new SAF scheme, more
sophisticated than the sequential one in order to improve the statistical
properties of the equivalent channel matrix.

\appendix
\subsection{Preliminaries}
\label{app:preliminaries}

For any linear fading Gaussian channel~
\begin{equation*}
  \my = \sqrt{\SNR}\,\He\,\mx + \mz
\end{equation*}%
where $\mz$ is an AWGN with $\EE\bigl\{\mz\transc{\mz}\bigr\}=\Id$ and
$\mx$ is subject to the input power constraint
$\text{Tr}\left\{\EE\left[\mx\transc{\mx}\right]\right\}\leq 1$, the
DMT $d_\mH(r)$ can be found as the exponent of the outage probability
in the high SNR regime, \ie,
\begin{align}
  \Pout(r\log\SNR) &\asympteq \textrm{Prob}\bigl\{\log\det\left(\Id+\SNR\,\He\transc{\He}\right)\leq r\log\SNR \bigr\} \nnb\\
  &= \textrm{Prob}\bigl\{\det\left(\Id+\SNR\,\He\transc{\He}\right)\leq \SNR^r \bigr\} \nnb\\
  &\asympteq \SNR^{-d_\mH(r)}. \label{eq:dmt}
\end{align}

\begin{lemma}[Calculation of diversity-multiplexing tradeoff]
  \label{lemma:cal-dmt}
  Consider a linear fading Gaussian channel defined by $\mH$ for which
  $\det\left(\Id+\SNR\,\He\transc{\He}\right))$ is a function of
  $\mlambda$, a vector of positive random variables. Then, the DMT
  $d_\mH(r)$ of this channel can be calculated as
  \begin{equation*}
    d_\mH(r) = \inf_{\Ocal(\malpha, r)} \varepsilon(\malpha)
  \end{equation*}%
  where $\malpha_i\defeq-\log v_i/\log\SNR$ is the exponent of $v_i$,
  $\Ocal(\malpha, r)$ is the outage event set in terms of $\malpha$
  and $r$ in the high SNR regime, and $\varepsilon(\malpha)$ is the
  exponential order of the pdf $p_{\malpha}(\malpha)$ of $\malpha$,
  \ie,
  \begin{equation*}
    p_{\malpha}(\malpha) \asympteq \SNR^{-\varepsilon(\malpha)}.
  \end{equation*}%
\end{lemma}%
\begin{proof}
  This lemma can be justified by \Eq{eq:dmt} using Laplace's method,
  as shown in \cite{Zheng_Tse}.
\end{proof}

\begin{lemma}\label{lemma:PDFexp}
  Let $X$ be a $\chi^2$-distributed random variable with $2t$ degrees
  of freedom and $Y$ be a uniformly distributed random variable in an
  interval including $0$. Define $\xi\defeq-\frac{\log X}{\log\SNR}$
  and $\eta\defeq-\frac{\log \Abssqr{Y}}{\log\SNR}$, then we have
  \begin{equation*}
    p_{\xi}\asympteq 
    \begin{cases}
      \SNR^{-\infty}& \text{for $\xi<0$,}\\
      \SNR^{-t\xi} & \text{for $\xi\geq0$;}
    \end{cases}%
  \end{equation*}%
  and 
  \begin{equation*}
    p_{\eta}\asympteq 
    \begin{cases}
      \SNR^{-\infty}& \text{for $\eta<0$,}\\
      \SNR^{-\eta/2} & \text{for $\eta\geq0$.}
    \end{cases}%
  \end{equation*}%
\end{lemma}

\begin{lemma} \label{lemma:bidiag}
  Let $\mG$ be $( k+1)\times(k+1)$ bidiagonal matrix defined by
\begin{equation*}
\mG\defeq x_0\,\Id + \matrix{\trans{\mbs{0}} & 0\\ \diag{(\mx)} & \mbs{0}}.
\end{equation*}
Then,
\begin{equation*}
\det\left(\Id+\mG\transc{\mG}\right) \geq \Abs{x_0}^{2(k+1)} +
\prod_{i=1}^k \left(1+\Abssqr{x_i}\right).  
\end{equation*}
\end{lemma}

%\subsection{Proof of Lemma~\ref{lemma:asymptgeq_D1}}
%\label{app:proof_asymptgeq_D1}
\begin{proof}
  Define $\mM_{k+1}\defeq\Id+\mG\transc{\mG}$ which is tridiagonal in
  the following form
\begin{equation*}
\matrix{
  1+\Abssqr{x_0} & x_0 x_1^* & \cdots & 0  \\
  x_0^* x_1 & 1+\Abssqr{x_0}+\Abssqr{x_1} & \ddots & \vdots \\
  \vdots &\ddots &\ddots & x_0 x_k^*  \\
  0 & \cdots & x_0^* x_k & 1+\Abssqr{x_0}+\Abssqr{x_k} }.
\end{equation*}%
For simplicity, let $X_i\defeq \Abssqr{x_i}$ for $i=0,\ldots,k$,
$D_k\defeq\det(\mM_{k})$ and use the formula for the calculation of
the determinant of a tridiagonal matrix~\cite{Horn}, we have
\begin{equation}
  \label{eq:recursive_D1}
  \begin{split}
    D_{k+1} &= (1+X_0+X_k)D_k - X_0 X_k D_{k-1} \\
            &= (1+X_0)D_k + X_k(D_k-X_0 D_{k-1}).
  \end{split}%
\end{equation}%
Let us rewrite the last equation as
\begin{equation}
  \label{eq:recursive_D2}
  D_{k+1} - X_0 D_k = X_k(D_k - X_0 D_{k-1}) + D_k
\end{equation}%
and define $B_k \defeq D_k - X_0 D_{k-1}$, from \Eq{eq:recursive_D1}
and \Eq{eq:recursive_D2}, we get
\begin{equation}
  \label{eq:recursive_vec}
  \matrix{D_{k+1}\\B_{k+1}} = \matrix{1+X_0 & X_k\\1 & X_k} \matrix{D_k\\B_k}.
\end{equation}%
First, it is easy to show that $D_2=X_0^2+2X_0+(X_1+1)$ and
$B_2=X_0+X_1+1$. Then, from \Eq{eq:recursive_vec}, it is obvious that,
as a polynomial of $(X_0,\ldots,X_k)$, $D_{k+1}$ has nonnegative
coefficients for any $k$. Finally, as a polynomial of $X_0$,
$D_{k+1}$'s coefficients can be found recursively using
\Eq{eq:recursive_D1} and we have
\begin{equation*}
  D_{k+1}(X_0) = X_0^{k+1} + \prod_{i=1}^k (1+X_i) + P(X_0).
\end{equation*}%
where $P(X_0)\geq0$ is a polynomial of $X_0$ and is always
nonnegative. Thus, we have 
\begin{equation*}
  D_{k+1} \geq X_0^{k+1} + \prod_{i=1}^k (1+X_i).
\end{equation*}%
%which can be used to get \Eq{eq:asymptgeq_D1}.\hfill \QED
\end{proof}

\subsection{Proof of Theorem~\ref{thm:theorem_UB}}
%\subsection{Proof of Lemma~\ref{lemma:asymptleq_D}}
\label{app:proof_theorem_UB}

The DMT of the genie-aided model can be obtained by considering the
equivalent channel matrix defined by \Eq{eq:GA-H}. First, it is
upper-bounded, as shown in the following lemma.
\begin{lemma}\label{lemma:asymptleq_D}
  For the genie-aided model \Eq{eq:GA-H}, let us define
  $\Abssqr{g_{\max}}\defeq\displaystyle\max_{\sss i=0\ldots N}
  \Abssqr{g_i}$, then we have
  \begin{equation}\label{eq:asymptleq_D}
    \begin{split}
      \det\left(\Id + \SNR\mH\transc{\mH}\right)
      &\asymptleq \left(1+\SNR\Abssqr{g_0}\right)^M \\
      &\quad + \left(1+\SNR \Abssqr{g_{\max}} \right)^{M-1}.
    \end{split}
  \end{equation}%
\end{lemma}%
\begin{proof}
  We can prove it in a recursive manner. First, any $(n+1)\times (n+1)$
  lower-triangular matrix $\mH_{n+1}$ can be written as
\begin{equation*}
\mH_{n+1}=
\matrix{
    \mH_n & \mbs{0} \\ \transc{\mv}_n & g} 
\end{equation*}
Let us define $D_{n+1}\defeq
\det\left(\Id+\SNR\mH_{n+1}\transc{\mH}_{n+1}\right)$ and $C\defeq
1+\SNR\Abssqr{g}$. Then, we have
\begin{equation*}
  \begin{split}
    D_{n+1} &= C\det\left(\Id + \frac{\SNR}{C}\mv_n\transc{\mv}_n+
      \SNR\transc{\mH}_n\mH_n\right) \\
    &\stackrel{(a)}{\leq} C\left(1 +\SNR\lambda_1 +
      \frac{\SNR}{C}{\Norm{\mv_n}}^2
    \right)\prod_{i=2}^n(1+\SNR\lambda_i)\\
    &=    C\,D_n + \SNR{\Norm{\mv_n}}^2\prod_{i=2}^n(1+\SNR\lambda_i)\\
    &\leq C\,D_n +
    \left(1+\SNR{\Norm{\mv_n}}^2\right)\left(1+\SNR\Frob{\mH_n}\right)^{n-1}\\
    &\leq C\,D_n +
    \left(1+\SNR\Frob{\mH_{n+1}}\right)^{n}
  \end{split}
\end{equation*}
%%     &\stackrel{(b)}{\leq} C\,D_n + \SNR{\Norm{\mv_n}}^2\left(\frac{1}{n-1}\sum_{i=2}^n(1+\SNR\lambda_i)\right)^{n-1}\\
with $\lambda_i$ the $i^\th$ smallest eigenvalue of
$\mH_n\transc{\mH}_n$. The inequality (a) comes from the fact that
$\mv_n\transc{\mv}_n$ has only one nonzero eigenvalue and that for any
nonnegative matrix $\mA$ and $\mB$, $\det(\mA+\mB)$ is maximized when
they are simultaneously diagonalizable and have eigenvalues in reverse
order. By setting $\mH_{n+1}=\mH$ of the genie-aided model, we have 
\begin{equation}\label{eq:ineq-recur}
    D_{n+1} \asymptleq C\,D_n + \left(1+\SNR \Abssqr{g_{\max}}\right)^n
\end{equation}%
since $\Frob{\mH_{n+1}} \leq M\Abssqr{g_0}+\sum_{j=1}^N
\Abssqr{g_j}\Frob{\mL_j}\asymptleq \Abssqr{g_{\max}}$ where we use the
fact that $\Frob{\mL_j}\asymptleq \SNR^0$ to meet the power constraint
\Eq{eq:PC}. The inequality \Eq{eq:ineq-recur} leads directly to
\Eq{eq:asymptleq_D} in a recursive manner.
\end{proof}

Then, the upper bound \Eq{eq:asymptleq_D} is achievable by setting
$\mH$ bidiagonal with
\begin{equation*}
\mH = g_0\,\Id + \matrix{\trans{\mbs{0}} & 0\\ g_{\max}\,\Id & \mbs{0}},
\end{equation*}
which can be justified by Lemma~\ref{lemma:bidiag}. This setting is
equivalent to using uniquely the relay with the best relay-destination
channel gain to send $\mx_{i-1}$ during the $i^\th$ slot. 

Now, define
$\malpha\defeq\left[\alpha_{g_0}\ldots\alpha_{g_N}\right]$, where
$\alpha_{g_i}$ is such that $\Abssqr{g_i}\asympteq
\SNR^{-\alpha_{g_i}}$. By applying Lemma~\ref{lemma:cal-dmt} on the
right hand side~(RHS) of \Eq{eq:asymptleq_D}, we get the DMT of the
genie-aided SAF
\begin{equation*}
  \bar{d}_\mH(r) = \inf_{\Ocal(\malpha, r)} \sum_{i=0}^N \alpha_{g_i}
\end{equation*}%
with 
\begin{equation*}
  \Ocal(\malpha, r) = \left\{{M \pstv{(1-\alpha_{g_0})}< r;}\atop{(M-1) \pstv{(1-\alpha_{g_i})}< r,\quad \text{for}\ i=1,\ldots,N}\right\}.
\end{equation*}%
Due to the symmetry of $\alpha_{g_i}$ for $i=1,\ldots,N$, we can solve
the linear programming problem by adding the constraint
$\alpha_{g_1}=\ldots=\alpha_{g_N}$. Applying
Theorem~\ref{thm:dmt-saf}, we can get the closed-form DMT
\Eq{eq:dmt_ub2}.

\subsection{Lower-bound on the DMT with Isolated Relays}
\label{app:proof_isolated}

\subsubsection{Dumb scheduling}
\label{sec:dump-scheduling}

In the $\tilde{N}=k\,N$ case with any integer $k$, a round-robin
scheme is optimal since the $\tilde{N}$ slots are equally protected by
all the relays. The RHS of \Eq{eq:asymptgeq_D1} becomes
\begin{equation}
  \label{eq:asymptgeq_D3}
  \left(1+\SNR\Abssqr{g_0}\right)^M 
       + \prod_{i=1}^{N} \left(1+\SNR
        \Abssqr{g_i\,h_i} \right)^k.
\end{equation}%
We carry out the same calculations as in section~\ref{sec:ub} with
some modifications. Define
$\malpha\defeq\left[\alpha_{g_0}\ldots\alpha_{g_N}\ 
  \alpha_{h_1}\ldots\alpha_{h_N}\right]$. By applying
Lemma~\ref{lemma:cal-dmt} on \Eq{eq:asymptgeq_D3}, we have
\begin{equation*}
  \underline{d}_\mH(r) = \inf_{\Ocal(\malpha, r)} \left(\alpha_{g_0} + \sum_{i=1}^N (\alpha_{g_i}+\alpha_{h_i})\right)
\end{equation*}%
with 
\begin{equation*}
  \Ocal(\malpha, r) = \left\{{M \pstv{(1-\alpha_{g_0})}< r;}\atop{k\sum_{i=1}^N \pstv{(1-\alpha_{g_i}-\alpha_{h_i})}< r}\right\}.
\end{equation*}%
Note that by using the variable changes $\alpha'_{g_i}\defeq
\alpha_{g_i}+\alpha_{h_i}$ for $i=1,\ldots,N$, we get a linear
programming problem with symmetry of
$\alpha'_{g_1},\ldots,\alpha'_{g_N}$. The optimum must satisfy
$\alpha'_{g_1}=\ldots=\alpha'_{g_N}=\beta$, and the optimization
problem reduces to
\begin{equation}\label{eq:tmp1}
  \underline{d}_\mH(r) = \inf_{\Ocal(\alpha_{g_0},\beta, r)} \left( \alpha_{g_0} + N\beta \right)
\end{equation}%
with 
\begin{equation*}
  \Ocal(\alpha_{g_0},\beta,r) = \left\{{M \pstv{(1-\alpha_{g_0})}< r;}\atop{(M-1)\beta < r}\right\}.
\end{equation*}%
Solving this problem, we get exactly \Eq{eq:dmt_ub2}.

In the $\tilde{N}= kN+m$ case, the RHS of \Eq{eq:asymptgeq_D1} is
directly revised as
%\begin{multline}
  \begin{equation}
  \label{eq:asymptgeq_D2}
%  \begin{split}
    \left(1+\SNR\Abssqr{g_0}\right)^M 
     + \left(\prod_{n=1}^N \left(1+\SNR \Abssqr{g_n\,h_n}
      \right)^k\right)\prod_{i=1}^m \left(1+\SNR \Abssqr{g_i\,h_i}
    \right).    
   \end{equation}
%  \end{multline}
%\end{equation}%
  Then, we have the same optimization problem \Eq{eq:tmp1} with
  different constraints, due to the relay ordering. Using the same
  variable changes, we have
\begin{equation*}
  \Ocal(\malpha, r) = \left\{{\D{M \pstv{(1-\alpha_{g_0})}< r;}\atop{\D k\sum_{i=1}^N \pstv{(1-\alpha'_{g_i})} + \sum_{i=1}^m \pstv{(1-\alpha'_{g_i})}< r;}}\atop{\max\{\alpha'_{g_1},\ldots,\alpha'_{g_m}\}\leq\min\{\alpha'_{g_{m+1}},\ldots,\alpha'_{g_N}\}}\right\},
\end{equation*}%
where the third constraint comes from the fact that $C_i \asympteq
\SNR^{1-\alpha'_{g_i}}$~($\SNR\Abssqr{b_i}\asympteq \SNR^0$). The second
and the third constraints together are equivalent to
\begin{multline}\label{eq:tmp2}
  \Biggl\{k\sum_{i=1}^N \pstv{(1-\alpha'_{g_i})} + \sum_{i=1}^m
  \pstv{(1-\alpha'_{g_{\Scal(i)}})} < r,\forall\ 
  \Scal\subseteq\{1,\ldots,N\}\ \text{and}\ \Abs{\Scal}=m \Biggr\},
\end{multline}%
from which we get a symmetric problem for $\alpha'_{g_i}$,
$i=1,\ldots,N$. We can then prove the same result as the previous
case.

\subsubsection{Smart scheduling}
\label{sec:smart-scheduling}

Using the two ``best'' relays, we can arrive at \Eq{eq:tmp2} with
$N=2$. Since our definition of ``best'' also corresponds to minimum
value of $\alpha'_{g_i}$, it is not difficult to verify that the
outage region in this case is included in the region~\Eq{eq:tmp2}.
Thus, the DMT is lower-bounded by that of the dumb scheduling and the
achievability is proved.
\hfill\QED

\subsection{Proof of Proposition~\ref{prop:prop1}}
\label{app:proof-2r3s}

\begin{fact}
Let $\mf \defeq \trans{\bigl[ f_1 \  f_2\bigr ]}$, 
$\mU \defeq \left[
\begin{IEEEeqnarraybox*}[\mysmallarraydecl]
[c]{,c/c,}
    u_{11} & 0 \\ u_{21} & u_{22}
\end{IEEEeqnarraybox*}\right]$
 and  $\He$ be a $3\times3$ upper-triangular matrix defined by
\begin{equation*}
  \He \defeq \matrix{\mU & \mbs{0} \\ \trans{\mf} & g}
\end{equation*}%
with $g$ being a scalar. Then, we have
\begin{equation}\label{eq:detM}
  \begin{split}
    \det(\Id + \SNR\,\He\transc{\He}) &= (1+\SNR\Abssqr{g})\det\bigl(\Id+\SNR\,\mU\transc{\mU}\bigr) \\
              &\quad + \SNR\Norm{\mf}^2 + \SNR^2\Abssqr{f_2 u_{11}} \\
              &\quad + \SNR^2\Abssqr{u_{22}f_1 - u_{21}f_2}.
  \end{split}
\end{equation}%
\end{fact}

Since non-zero multiplicative constants independent of SNR do not
appear in the high SNR regime analysis, from \Eq{eq:sequential-H}, we
consider the following matrix 
\begin{equation}\label{eq:H2}
  \mH = \matrix{g_0 & 0 & 0 \\ g_1\,h_1 & g_0 & 0 \\ g_2\,\gamma_{12}\,h_1 & g_2\,h_2 & g_0},
\end{equation}
where the coefficients $\sqrt{\SNR}\,b_1$ and $\sqrt{\SNR}\,b_2$ are
neglected~($\SNR\Abssqr{b_i}\asympteq \SNR^0$). With \Eq{eq:detM}, we
can now obtain the outage event set, in terms of the entries of $\mH$.

In order to apply lemma~\ref{lemma:cal-dmt}, however, we must get the
outage event set in the high SNR regime, in terms of $\alpha$. To this
end, we must rewrite $\Abssqr{u_{22}f_1-u_{21}f_2}$ in \Eq{eq:detM} in
a more convenient form of positive variables. Let us use the notation
$V = \Abssqr{v}$ for $v$ being any variable. Then, from \Eq{eq:detM}
and \Eq{eq:H2}, we have
\begin{align*}
  F_1 &\asympteq G_2H_1\Gamma_{12};&  F_2&\asympteq G_2H_2; \\
  U_{11} &\asympteq U_{22} \asympteq G_0;& U_{21} &\asympteq G_1H_1.
\end{align*}%
Let us rewrite 
\begin{equation*}
  \begin{split}
    \Abssqr{u_{22}f_1-u_{21}f_2} &= U_{22}F_1 + U_{21}F_2
 - 2\sqrt{U_{21}U_{22}F_1 F_2} \cos{\theta} \\
      &= (1-\cos{\theta})(U_{22}F_1 + U_{21}F_2) \\
      & \quad + \cos{\theta} \Abssqr{\sqrt{U_{22}F_1}-\sqrt{U_{21}F_2}}
  \end{split}
\end{equation*}%
with $\theta$ uniformly distributed in $[0,\pi]$ and is independent of
the other random variables. The outage probability conditioned on
$\theta$ is maximized when $\theta$ is close to $\pstv{0}$, where
$1-\cos{\theta}\approx \frac{\theta^2}{2}$. In this region, we have 
\begin{equation}
  \begin{split}
    \Abssqr{u_{22}f_1-u_{21}f_2} &\asympteq \frac{\theta^2}{2}(U_{22}F_1 +
    U_{21}F_2) \\
    &\quad + \Abssqr{\sqrt{U_{22}F_1}-\sqrt{U_{21}F_2}} \label{eq:misc1}
  \end{split}
\end{equation}%
Then, from \Eq{eq:detM} and \Eq{eq:misc1}, we have the outage region
$\Ocal(\He,r)$
\begin{equation}
  \left\{
      \begin{array}{rcl}
        (1+\SNR G_0)\det(\Id+\SNR\,\mU\transc{\mU}) &\asymptleq& \SNR^r \\
        1+\SNR (F_1 + F_2) &\asymptleq& \SNR^r \\
        1+\SNR^2 F_2 U_{11} &\asymptleq& \SNR^r \\
        1+\SNR^2 \theta^2 (U_{22}F_1+U_{21}F_2) &\asymptleq& \SNR^r \\
        1+\SNR^2 \Abssqr{\sqrt{U_{22}F_1}-\sqrt{U_{21}F_2}} &\asymptleq& \SNR^r       \end{array}
\right\} \label{eq:misc2}
\end{equation}%
The last inequality in \Eq{eq:misc2} implies  
\begin{equation*}
  1+\SNR^2 (U_{22}F_1+U_{21}F_2) \asymptleq \SNR^r + 2\SNR^2\sqrt{U_{21}U_{22}F_1F_2},
\end{equation*}
which means that, in the high SNR regime, the outage region
$\Ocal(\He,r)$ is included\footnote{In this case, we have
  $\Ocal(\He,r)\subseteq\Ocal(\malpha,r)$ but
  $\Ocal(\malpha,r)\nsubseteq\Ocal(\He,r)$} in the region
$\Ocal(\malpha,r)$ defined by %% with $\malpha\defeq[\alpha_{g_0}\ \alpha_{g_1}\ 
%% \alpha_{g_2}\ \alpha_{h_1}\ \alpha_{h_2}\ \alpha_{t}\ 
%% \alpha_{\theta}]$
\begin{equation*}
  \left\{
      \begin{array}{rcl}
        3(1-\alpha_{g_0}) &\leq& r\\
        (1-\alpha_{g_0}) + (1-\alpha_{g_1}-\alpha_{h_1}) &\leq& r\\
        2-\alpha_{g_0}-\alpha_{g_2}-\alpha_{h_2} &\le& r\\
        1 - \alpha_{g_2} - \alpha_{\gamma_{12}} - \alpha_{h_1} &\leq& r\\
        2-\alpha_{g_0}-\alpha_{g_2}-\alpha_{\gamma_{12}}-\alpha_{h_1} -\alpha_{\theta}&\leq&r \\
        2-\alpha_{g_1}-\alpha_{g_2}-\alpha_{h_1}-\alpha_{h_2}  -\alpha_{\theta}&\leq&r \\
        2-\alpha_{g_0}-\alpha_{g_2}-\alpha_{\gamma_{12}}-\alpha_{h_1} &\le& \max\left\{r, \phi(\malpha)\right\}\\
        2-\alpha_{g_1}-\alpha_{g_2}-\alpha_{h_1}-\alpha_{h_2} &\le& \max\left\{r, \phi(\malpha) \right\}\\
      \end{array} \right\}
\end{equation*}%
with $\phi(\malpha)\defeq
2-\frac{1}{2}(\alpha_{g_0}+\alpha_{g_1}+\alpha_{\gamma_{12}}+\alpha_{h_2})-\alpha_{h_1}-\alpha_{g_2}$. 
Let us define
\begin{eqnarray*}
 \Ocal_{\Tcal}(\malpha,r) &\defeq& \Ocal(\malpha,r)\cap \Tcal(\malpha,r)\\
 \Ocal_{\overline{\Tcal}}(\malpha,r) &\defeq& \Ocal(\malpha,r)\cap \overline{\Tcal}(\malpha,r)
\end{eqnarray*}
with 
\begin{equation*}
  \Tcal(\malpha,r) \defeq \left\{\malpha~:\quad r\leq\phi(\malpha)\right\}.
\end{equation*}%
Since $\Ocal(\malpha, r)=\Ocal_{\Tcal}(\malpha, r)\cup\Ocal_{\overline{\Tcal}}(\malpha, r)$, we have
\begin{equation*}
  \inf_{\Ocal(\malpha, r)} \varepsilon(\malpha) = \min\left\{\inf_{\Ocal_{\Tcal}(\malpha, r)} \varepsilon(\malpha), \inf_{\Ocal_{\overline{\Tcal}}(\malpha, r)} \varepsilon(\malpha)\right\},
\end{equation*}%
with
$\varepsilon(\malpha)=\alpha_{g_0}+\alpha_{g_1}+\alpha_{g_2}+\alpha_{h_1}+\alpha_{h_2}+\alpha_{\gamma_{12}}+\frac{1}{2}\alpha_{\theta}$
by lemma~\ref{lemma:PDFexp} and the independence between the random
variables. Thus, the DMT can be obtained with two linear
optimizations. This problem can be solved numerically using
sophisticated linear programming algorithms or softwares. If the relay
ordering is such that $\Abs{h_2}>\Abs{h_1}$, we add
$\alpha_{h_1}>\alpha_{h_2}$ to the constraints and carry out the same
optimization problem. We can finally get the DMTs of
Fig.~\ref{fig:dmt-proposed}.

\hfill\QED

\end{document}